\documentclass[aps,twocolumn,preprintnumbers,amsmath,amssymb,longbibliography]{revtex4-1}

\usepackage{graphicx}
\usepackage{dcolumn}
\usepackage{bm}
\usepackage{multirow}
\usepackage{color}
\usepackage{amsmath}
\usepackage{gensymb}
\usepackage[colorlinks=true]{hyperref}

\begin{document}
\title{
In-plane Propagation of Light in Transition Metal Dichalcogenide Monolayers: Optical Selection Rules 
}

\author{G.~Wang$^1$}
\author{C.~Robert$^1$}
\author{M.~M.~Glazov$^{2}$}
\author{F.~Cadiz$^1$}
\author{E.~Courtade$^1$}
\author{T.~Amand$^1$}
\author{D.~Lagarde $^1$}
\author{T. Taniguchi$^3$}
\author{K. Watanabe$^3$}
\author{B.~Urbaszek$^1$}
\author{X.~Marie$^1$}

\affiliation{%
$^1$Universit\'e de Toulouse, INSA-CNRS-UPS, LPCNO, 135 Av. Rangueil, 31077 Toulouse, France}
\affiliation{$^2$Ioffe Institute, 194021 St.\,Petersburg, Russia}
\affiliation{$^3$National Institute for Materials Science, Tsukuba, Ibaraki 305-0044, Japan}

\begin{abstract}
The optical selection rules for inter-band transitions in WSe$_2$, WS$_2$ and MoSe$_2$ transition metal dichalcogenide monolayers are investigated by polarization-resolved photoluminescence experiments with a signal collection from the sample edge.
These measurements reveal a strong polarization-dependence of the emission lines. We see clear signatures of the emitted light with the electric field oriented perpendicular to the monolayer plane, corresponding to an inter-band optical transition forbidden at normal incidence used in standard optical spectroscopy measurements. The experimental results are in agreement with the optical selection rules deduced from group theory analysis, highlighting the key role played by the different symmetries of the conduction and valence bands split by the spin-orbit interaction. These studies yield a direct determination on the bright-dark exciton splitting, for which we measure 40 $\pm1$ meV and 55 $\pm2$ meV  for WSe$_2$ and WS$_2$ monolayer, respectively.

\end{abstract}


\maketitle

Two-dimensional crystals of transition metal dichalcogenides such as MX$_2$ (M=Mo, W; X=S, Se, Te) are promising atomically flat semiconductors for applications in electronics and optoelectronics \cite{Butler:2013a,Geim:2013a,Mak:2010a, Splendiani:2010a, Wang:2012c}. The optical properties of transition metal dichalcogenides (TMD) monolayers (MLs) are governed by very robust excitons with binding energy of the order of 500 meV \cite{He:2014a,Ugeda:2014a,Chernikov:2014a,Ye:2014a,Qiu:2013a,Ramasubramaniam:2012a,Wang:2015b}. The interplay between inversion symmetry breaking and strong spin-orbit coupling in these MLs also yields unique spin/valley properties \cite{Xiao:2012a,Sallen:2012a,Mak:2012a,Kioseoglou:2012a,Cao:2012a,Jones:2013a,Yang:2015a}.
Due to the two-dimensional (2D) character of the layered materials, the band-to-band transitions are predicted to be anisotropic for light propagating parallel to the plane of the ML. The insights gained from these type of experiments in III-V semiconductor quantum wells \cite{Miller:1985,Marzin:1985} were crucial for designing optoelectronic devices. For 2D materials based on TMDs the light polarized perpendicular to the ML ($z$ direction) should involve transitions with energies different from the transitions observed for in-plane polarized light \cite{Glazov:2014a,Echeverry:2016}. So far, however, optical spectroscopy measurements in TMD MLs have only been made for normal incidence for which the $z$ polarization is not accessible. Indeed the natural geometry in optical spectroscopy consists in exciting and collecting light from the top of the sample, with light wave-vectors perpendicular to the ML plane.  In addition to the measurement of the predicted anisotropy of the interaction of light, optical experiments performed for light propagating parallel to the ML should bring precious information on the detailed band structure of these 2D materials. In particular, it allows for a straightforward determination of the energy difference between bright and dark excitons, for which a direct measurement is still lacking \cite{Zhang:2015d,Wang:2015e,Arora:2015a,Withers:2015}.

In this Letter we present the first measurements of the luminescence properties of TMD MLs for light propagating along the plane of the layer. In this geometry the electric field of the optical radiation can be either parallel or perpendicular to the ML. We measure the polarization-dependent emission properties of different TMD MLs for this in-plane optical excitation and detection geometry. For WSe$_2$ and WS$_2$ MLs, a new luminescence line emerges for the polarization perpendicular to the 2D material plane, corresponding to a $z$-dipole transition. This transition is forbidden at normal incidence in linear optical spectroscopy because the electromagnetic field is transverse. Our measurements yield a direct determination of the bright-dark exciton splitting, in agreement with the selection rules deduced from group theory. We find 40 and 55 meV in WSe$_2$ and WS$_2$ MLs, respectively. For MoSe$_2$ ML, no signature of the  $"$dark$"$  state is evidenced as a consequence of the very low population of the dark exciton states which lie at higher energy compared to the bright ones \cite{Kosmider:2013b, Echeverry:2016,Kormanyos:2015a,Dery:2015a}.  

\begin{figure*}
\includegraphics[width=0.8\textwidth,keepaspectratio=true]{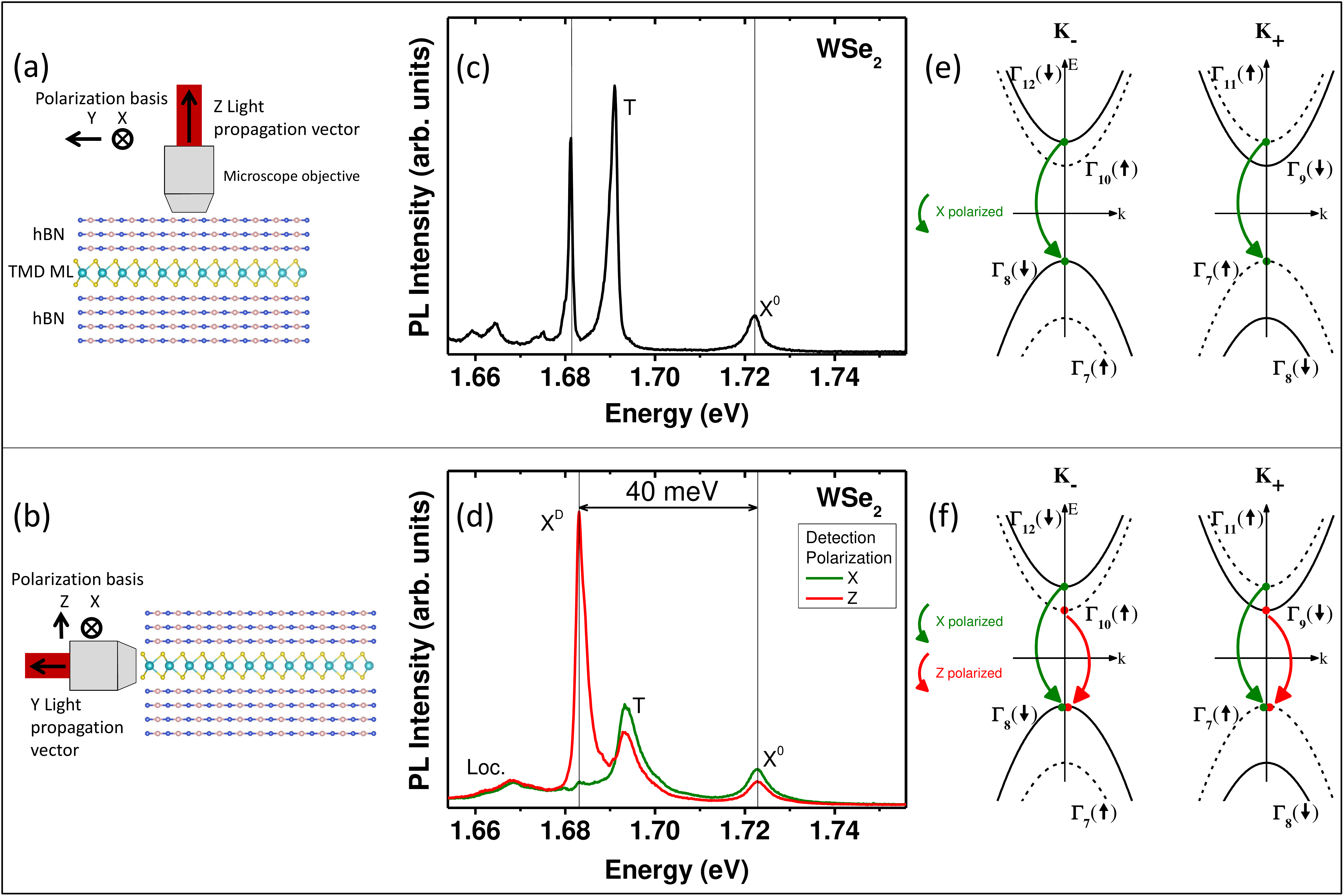}
\caption{ Schematics of the excitation/detection geometry of the PL for (a) light propagating perpendicular to the ML plane \& detection of the PL from the top of the sample,  (b) light propagating parallel to the ML plane \& detection of the PL from the edge of the sample. (c) Detection of the PL from the top of the sample. The PL spectrum of  hBN/WSe$_2$ ML /hBN at $T=13$~K; the polarization of the excitation/detected light is in the ML plane ($x$-direction). (d) Detection of the PL from the edge of the sample. The PL spectrum of  hBN/WSe$_2$ ML /hBN at $T=13$~K; the polarization of the detected light is in the ML plane ($x$-direction), green line, or perpendicular to it ($z$-direction), red line. (e) Sketch of the band structure of WSe$_2$  ML. The bands are labeled in each valley by the corresponding irreducible spinor representations with arrows in parentheses indicating the dominant electron spin orientation. The green arrows show the transitions optically active for the $x$-polarized light,  (f) the green and red arrows show the transitions optically active for $x$-polarized and $z$-polarized light, respectively (light propagating parallel to the ML plane).}\label{fig:r1} 
\end{figure*}

We have investigated MX$_2$ MLs encapsulated in hexagonal boron nitride (hBN) and transferred onto an SiO$_2$(90 nm)/Si substrate, see schematics in Fig.~\ref{fig:r1}(a,b). The design of these samples is critical to the success of the experiment. It was shown recently that encapsulation of MX$_2$ ML in high quality hBN results in very high quality samples where surface protection and substrate flatness yield very small photoluminescence (PL) or reflectivity linewidths, in the range $2-5$ meV at low temperature \cite{Wang:2017b,Ajayi:2017,Cadiz:2017a, Manca:2017a}. 
In the present investigation the narrow exciton lines will allow us to identify clearly transitions involving different bands for different polarizations of the light propagating in the ML plane. These van der Waals heterostructures are obtained by mechanical exfoliation of bulk MX$_2$ (from 2D Semiconductors, USA) and hBN crystals \cite{Taniguchi:2007a}, following the fabrication technique detailed in Ref. \cite{Cadiz:2017a}. 
The typical thickness of the hBN layers is $\sim10$~nm and the in-plane size of the MX$_2$ ML is  $\sim 10\times10$~$\mu$m$^2$. The samples are held on a cold finger in a closed-cycle cryostat. Two configurations for the microscope objective inside the cryostat are used for the excitation and collection of PL along or perpendicular to the ML plane, Fig.~\ref{fig:r1}(a,b). Attocube X-Y-Z piezo-motors allow for positioning with nm resolution of the ML with respect to the microscope objective (numerical aperture NA$=0.82$) used for excitation and collection of luminescence.
The ML is excited by a continuous wave green laser (2.33 eV). For WSe$_2$ and MoSe$_2$ MLs similar results have been obtained with He-Ne laser excitation (1.96 eV). The laser average power is about 50~$\mu$W. The excitation laser and detection spot diameter is $\sim{1}\mu$m. The PL signal is dispersed in a spectrometer and detected with a Si-CCD camera \cite{Wang:2014b}.
For the measurements from the edge of the sample, the ratio between the focused laser spot diameter and the thickness of the ML is smaller than 1000. Though challenging from the point of view of the required alignment accuracy this experiment can be successful as shown below thanks to (i) the very large absorption coefficient of the TMD ML for in-plane polarized light \cite{Li:2014b}, (ii)~the longer interaction length between the light and 2D material compared to normal incidence excitation configuration and (iii) the detection efficiency of our set-up, designed for studies of single photon emitters~\cite{Belhadj:2009}. 

Figure~\ref{fig:r1}(c) presents the PL spectrum at $T=13$~K of the WSe$_2$ ML in the standard configuration, i.e. propagation of light perpendicular to the ML. We observe clearly the peaks corresponding to the recombination of neutral exciton X$^0$ ($1.722$~eV), trion -- charged exciton -- T ($1.690$~eV) and lower energy lines ($1.65 - 1.68$~eV) usually attributed to localized excitons, in agreement with already published results \cite{Jones:2015a,Arora:2015a,Wang:2015e}. For this geometry where the light is polarized in the ML plane, the detected neutral exciton luminescence X$^0$ corresponds to the radiative recombination involving both $\Gamma_{11}$ conduction band (CB) and $\Gamma_7$ valence band (VB) in the valley $\bm K_+$ and $\Gamma_{12}$ conduction band and $\Gamma_8$ valence band in the valley $\bm K_-$, see the green arrows in Fig.~\ref{fig:r1}(e). Both transitions conserve the spin. In contrast the transitions in the $\bm K_+$-valley between the $\Gamma_9$ CB and $\Gamma_7$ VB with opposite spins ($\Gamma_{10}$ CB and $\Gamma_{8}$  VB in the $\bm K_-$-valley) are optically forbidden for the in-plane polarized light. The energy difference between the corresponding dark exciton and the bright X$^0$ depends both on the spin-orbit splitting in the conduction band $\Delta_{SO}^{CB}$  and the short range part of the electron-hole exchange interaction, where $\Delta_{SO}^{CB}$ is the energy difference between $\Gamma_9$ ($\Gamma_{10}$) CB  and $\Gamma_{11}$ ($\Gamma_{12}$) in valley ${\bm K}_+$ {($\bm K_-$)}. The calculations predict values $\Delta_{\rm Bright-Dark}$ of a  few tens of meV for WSe$_2$ ML~\cite { Echeverry:2016,Qiu:2015a}.

\begin{figure}
\includegraphics[width=0.4\textwidth,keepaspectratio=true]{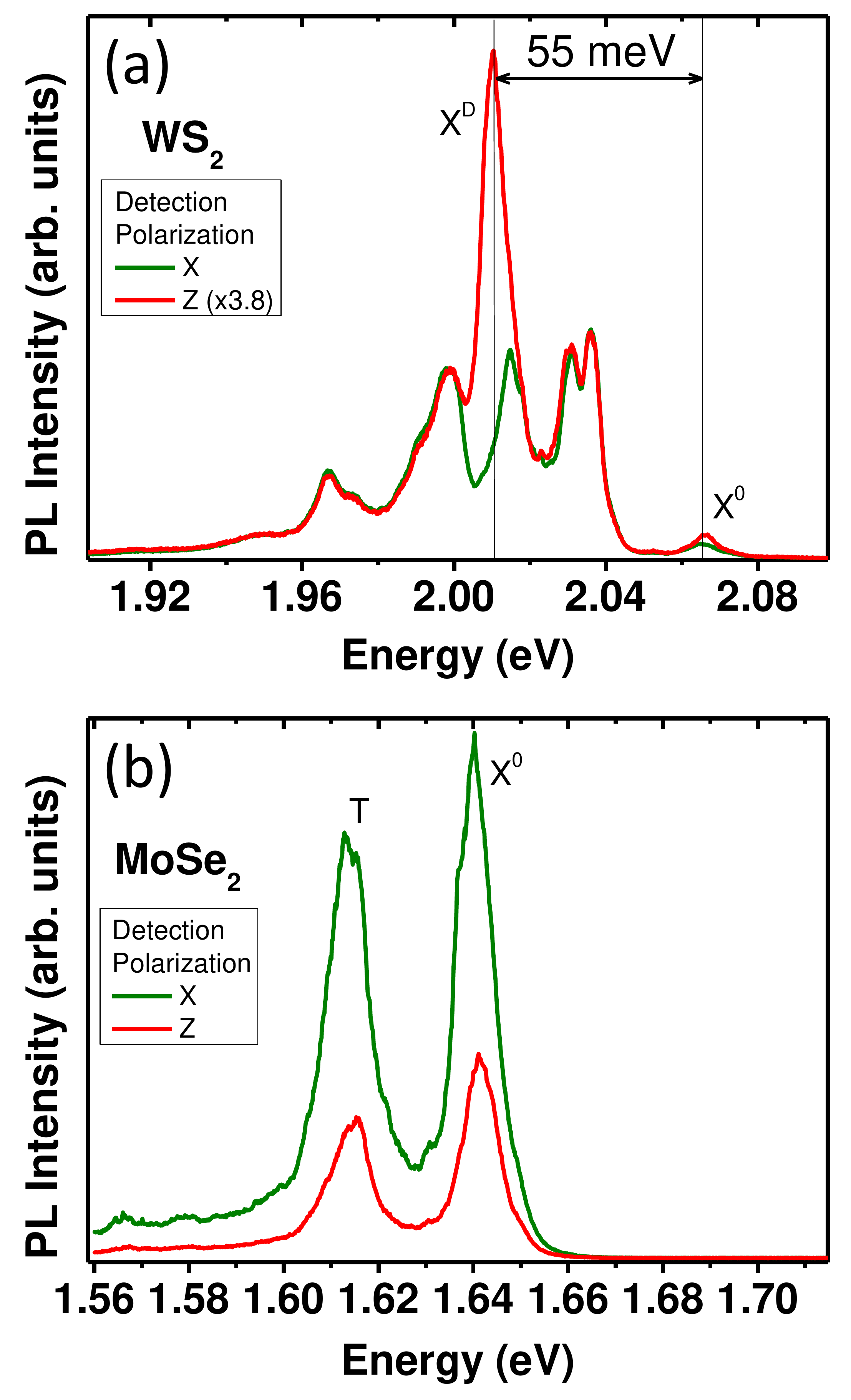}
\caption{Detection of the PL from the edge of the sample. The PL spectrum  at $T=13$~K for (a)  hBN/WS$_2$ ML /hBN and (b) hBN/MoSe$_2$ ML /hBN; the polarization of the excitation/detected light is in the ML plane ($x$-direction), green line, or perpendicular to it ($z$-direction), red line.}\label{fig:r2} 
\end{figure}

We present now the key results associated to the measurements from the edge of the sample, enabling the excitation and collection of the PL signal emitted with a wave-vector parallel to the ML, as schematically depicted in Fig.~\ref{fig:r1}(b). The great advantage of this geometry is that it is suitable for measuring the interaction of the 2D material with light for both polarization directions, parallel to the plane of the ML -- $x$-polarized -- as in Fig.~\ref{fig:r1}(c) or perpendicular to it -- $z$-polarized. The optical selection rules, which depend intimately on the band structure of the ML and the exciton symmetry, can thus be revealed. In Fig.~\ref{fig:r1}(d) the PL spectra of the WSe$_2$ ML for both in-plane ($x$) and perpendicular ($z$) to the plane polarization are displayed. As expected the in-plane polarized PL spectrum (green line) is very similar to the normal incidence excitation/detection geometry shown in Fig.~\ref{fig:r1}(c). As the polarization of the detected luminescence is identical, the same optical selection rules apply and we observe the three lines associated to neutral (X$^0$), charged (T) and localized excitons. 
The detection energies of the lines are identical for both geometries but we note a larger broadening for excitation/detection from the sample edge resulting probably from the longer interaction length in the 2D material. Remarkably a new line labelled X$^D$ shows up in addition to the previous ones when the luminescence polarized perpendicular to the ML is detected, red line in Fig.~\ref{fig:r1}(d). In agreement with the selection rules detailed below, this peak corresponds to the radiative recombination of excitons involving the transitions between the bottom $\Gamma_9$ {($\Gamma_{10}$)} CB and topmost $\Gamma_7$ {($\Gamma_8$)} VB in the valley $\bm K_+$ ($\bm K_-$). We recall that these transitions are optically forbidden (``dark excitons'') for in-plane polarized light, in agreement with the $x$-polarized spectrum in Fig.~\ref{fig:r1}(d).   As a consequence the energy difference between X$^0$ and X$^D$ in Fig.~\ref{fig:r1}(d) is a direct measurement of the bright-dark exciton splitting. We find $\Delta_{\rm Bright-Dark}=40\pm1$~meV in WSe$_2$.
Figure~\ref{fig:r2}(a) presents the result of the same experiment performed on WS$_2$ ML at $T=13$~K. Again when the luminescence polarized perpendicular to the ML is detected, a new line (X$^D$) shows up and we measure a bright-dark exciton splitting energy of 55 $\pm2$ meV. These accurate measurements will be a key element to improve the parametrization of ab-initio calculations of the band structure \cite{Echeverry:2016}.

Finally we perform the same investigation for a MoSe$_2$ ML both at $T=13$~K and 300 K. No spectral signature of the X$^D$ line is observed  in MoSe$_2$ ML in Fig.~\ref{fig:r2}(b) at low temperature and in Fig.~S4 of the supplement~\cite{suppl1} at higher temperature. This behavior is compatible with our current understanding of the MoSe$_2$ ML band structure. Though the exact value of $\Delta_{\rm Bright-Dark}$ is still unknown, both theoretical and experimental investigations predict that the dark exciton lie at higher energy compared to the bright one (in contrast to WSe$_2$ and WS$_2$ ML) \cite{Zhang:2015d, Wang:2015e,Arora:2015a,Withers:2015}. The main reason is the change of sign of the spin-orbit splitting in the conduction band between these different materials \cite{Kosmider:2013b}. Very recent investigations based on the coupling of the excitons to transverse high magnetic fields (tens of teslas) \cite{Zhang:2017a,Molas:2017},  or surface plasmon polaritons \cite{Zhou:2017} inferred similar bright-dark exciton splitting energies. We emphasize that our measurements are based on the intrinsic properties of the MLs and do not require any external perturbation in addition to the light.
For the sake of completeness we have also investigated the dependence of the PL spectra detected from the edge of the sample as a function of the  polarization of the excitation laser propagating along the ML plane. In the previous experiments, Figs.~\ref{fig:r1}, \ref{fig:r2}, the laser was polarized in the ML plane along $x$-axis. As expected, the PL intensity for the $z$-polarized excitation laser is reduced (see Fig.~S2 in \cite{suppl1}). However the relative oscillator strength of the $x$-polarized and $z$-polarized transition at a given energy can not be deduced from these measurements because of the non-resonant energy of the excitation laser where the selection rules are difficult to determine.

The conclusions drawn from the experimental results are confirmed by the group theory analysis of the selection rules.
Let us recall that the orbital Bloch functions of the valence band in $\bm K_+$ and $\bm K_-$ transform according to the same scalar representation $\Gamma_1$ (notations of Ref.~\cite{Koster:1963a}) of $C_{3h}$ point group relevant at the $\bm K_\pm$ points. The conduction band orbital Bloch functions transform according to $\Gamma_2$ and $\Gamma_3$ respectively. As a result, the spinor representations for the valence band are $\Gamma_7$ ($\uparrow$) and $\Gamma_8$ ($\downarrow$), while for the conduction band, Fig.~\ref{fig:r1}(e), they are
\begin{subequations}
\label{products}
\begin{align}
\Gamma_2 \times \Gamma_7 = \Gamma_{11}, \quad \Gamma_2 \times \Gamma_8 = \Gamma_9,\\
\Gamma_3 \times \Gamma_7 = \Gamma_{10}, \quad \Gamma_3 \times \Gamma_8 = \Gamma_{12}
\end{align}
\end{subequations}
Interband optical excitation gives rise to the electron-hole pairs bound into excitons by the Coulomb interaction. The excitonic states with $1s$ envelope function detected in our experiments transform according to the representations ${\Gamma_{\rm X}} = \Gamma_c\times\Gamma_{v}^*$, where $\Gamma_c$ is the representation of the conduction band state and $\Gamma_v$ is the representation of the empty valence band state. Note that the hole state is the time-reversed of the unoccupied valence band state therefore the conjugation of $\Gamma_v$ is needed~\cite{Glazov:2015a}. The exciton state is optically active in a given polarization if $\Gamma_{\rm X}$ contains the irreducible representation according to which the corresponding polarization vector $\bm e$ transforms.  As a consequence for the optically active excitons $\Gamma_X$ must contain $\Gamma_2+\Gamma_3$, i.e., the in-plane polarization, or $\Gamma_4$, $z$-polarization. Hereafter we consider only vertical optical transitions and obtain for possible symmetries of the exciton
 \begin{subequations}
\label{selection:prods}
\begin{align}
& \Gamma_{12} \times \Gamma_8^* = \Gamma_3, \quad & \Gamma_{11} \times \Gamma_7^*=\Gamma_2 \label{circ}\\
&\Gamma_{10} \times \Gamma_{8}^* = \Gamma_4, \quad & \Gamma_{9} \times \Gamma_7^* = \Gamma_4.
\end{align}
\end{subequations}
The basic functions of $\Gamma_2$ and $\Gamma_3$ irreducible representations transform as $x\pm \mathrm i y$, respectively. Hence, Eq.~\eqref{circ} describes the   excitons active in the $\sigma^+$ and $\sigma^-$ polarizations at the normal incidence of radiation, see green arrows in Fig.~\ref{fig:r1}(e,f). Here the electron spin in the course of the interband transition is conserved and the polarization of the PL is determined by the orbital character of the Bloch functions. By contrast, the transitions involving opposite spins for the conduction and valence band states are forbidden at the normal incidence because they couple with the light of $z$-polarization, i.e., normal to the ML plane. These transitions are depicted by red arrows in Fig.~\ref{fig:r1}(f). This is in agreement with the measurement of the additional line X$^D$ which is $z$-polarized in Figs.~\ref{fig:r1}(d) and \ref{fig:r2}(a) for WSe$_2$ and WS$_2$ ML respectively.  

A deeper insight into the symmetry of excitonic states can be obtained by considering the irreducible representations of the $D_{3h}$ point group relevant for the overall symmetry of the ML. Such an analysis allows one to study the mixing of excitons in different valleys. The results in the supplement~\cite{suppl1} demonstrate that out of two $z$-polarized states, $\Gamma_{X_1} = \Gamma_{10} \times \Gamma_{8}^*$ and $\Gamma_{X_2} =\Gamma_{9} \times \Gamma_7^*$ one linear combination with equal weights is active in the $z$-polarization and can be attributed to the X$^D$ line, while another one is forbidden.

As these transitions require spin-orbit interaction to induce spin mixing, their oscillator strengths are expected to be much weaker than the one for in-plane polarized light.  It can be qualitatively described by taking into account  the interaction of the upper valence bands $\Gamma_8$ and $\Gamma_7$ with remote bands with different orbital character and opposite spin orientations, see supplement~\cite{suppl1}. Recent density functional theory calculations performed at the GW level combined with Bethe-Salpeter equation for excitons predict that the  out-of plane contribution is $\sim 10^3$ times smaller than the in-plane one \cite{Echeverry:2016}. Though our experiments clearly evidence both transitions, the measured PL intensities cannot be used for the accurate determination of the relative ratio of the oscillator strengths since the non-resonant excitation results in an unknown populations of the exciton states X$^0$ and X$^D$. Particularly, the assumption of a thermodynamical equilibrium between the X$^0$ and X$^D$ is questionable considering the very short radiative lifetime of X$^0$ \cite{Korn:2011a,Lagarde:2014a,Robert:2016a}.  Further investigations based, e.g., on absorption or photocurrent measurements performed with strictly resonant excitation are required.

 Finally we note in Fig.~\ref{fig:r1}(c) a PL line at the energy X$^D$  (1.68 eV) for normal incidence excitation/detection where the optical selection rules dictate that this transition should be optically forbidden. The observation of this line is, actually, not surprising: First it can have a purely geometric origin.
As we use a microscope objective with high NA, the electric field vector at the focal tail has a significant component along the $z$-axis which enables excitation/detection of the X$^D$ transition even at the normal incidence. The percentage of the $z$-mode to the total intensity was estimated to be $\sim 9\%$ in a similar study performed in GaAs quantum wells \cite{Schardt:2006}. Second, a lowering of the symmetry of the 2D crystal due to local strain or ripples can induce a small mixing between bright and dark excitons, yielding the observation of the X$^D$ component even for normal incidence~\cite{Bayer:2002}. The same arguments can explain the observation of both X$^0$ and T lines for the out-of-plane polarization in Figs.~\ref{fig:r1}(d), \ref{fig:r2}, where the luminescence is collected from the edge of the ML.
To elucidate the origin we note that by detecting only the central part of the PL spot for normal incidence excitation/detection, the X$^D$ line totally vanishes confirming that the $z$-component is located off-axis only, see Fig.~S5 of supplement~\cite{suppl1}. This demonstrates that the X$^D$ line observed in Fig.~\ref{fig:r1}(c) is simply linked to the geometry of the experiment based on a microscope objective with a large numerical aperture. 
  
In conclusion the first measurements of the optical properties of transition metal dichalcogenide MLs for light propagating along the plane are reported and discussed in terms of optically allowed and forbidden transitions depending on the light polarization and propagation direction. These experiments reveal important features of the band structure of these atomically-thin semiconductors. In addition to their importance for the knowledge of the band structure and excitonic properties in TMDC MLs, these experiments pave the way to the investigation of  waveguides heterostructures and devices based on 2D materials.

\emph{Acknowledgements.}  
We thank ERC Grant No. 306719, ITN Spin-NANO Marie Sklodowska-Curie grant agreement No 676108, ANR MoS2ValleyControl, Programme Investissements d Avenir ANR-11-IDEX-0002-02, reference ANR-10- LABX-0037-NEXT for financial support and Laboratoire International Associe ILNACS CNRS-Ioffe. X.M. also acknowledges the Institut Universitaire de France. K.W. and T.T. acknowledge support from the Elemental Strategy Initiative conducted by the MEXT, Japan and JSPS KAKENHI Grant Numbers JP26248061, JP15K21722 and JP25106006. M.M.G. is grateful to the RFBR, Dynasty Foundation and RF President grant MD-1555.2017.2 for partial support.


%

\renewcommand{\thefigure}{S\arabic{figure}}
\renewcommand{\thesection}{S\Roman{section}}
\setcounter{figure}{0}   
\section{Supplemental material}

\subsection{Extended group theoretical analysis}

The analysis presented in the main text is based on the representations of the $C_{3h}$ point symmetry group relevant for the single valley of the TMD ML. It is instructive to analyse the symmetry of the excitonic states in terms of irreducible representations of the $D_{3h}$ point symmetry group relevant for the TMD ML as a whole. Making use of Ref.~\cite{Koster:1963a} we establish the following compatibility of the irreducible representations of these two point groups
\begin{subequations}
\label{corresp}
\begin{align}
\Gamma_5^{D_{3h}} \to \Gamma_5^{C_{3h}} + \Gamma_6^{C_{3h}},\\
\Gamma_7^{D_{3h}} \to \Gamma_7^{C_{3h}} + \Gamma_8^{C_{3h}},\\
\Gamma_8^{D_{3h}} \to \Gamma_9^{C_{3h}} + \Gamma_{10}^{C_{3h}},\\
\Gamma_9^{D_{3h}} \to \Gamma_{11}^{C_{3h}} + \Gamma_{12}^{C_{3h}}.
\end{align}
\end{subequations}
Note that the different irreducible representations relevant for the Kramers-degenerate states from the valleys $\bm K_+$ and $\bm K_-$ join in the $D_{3h}$ point symmetry group. Note that these representations are self-conjugate.

\begin{figure}[htbp]
\includegraphics [width=0.7\linewidth,keepaspectratio=true]{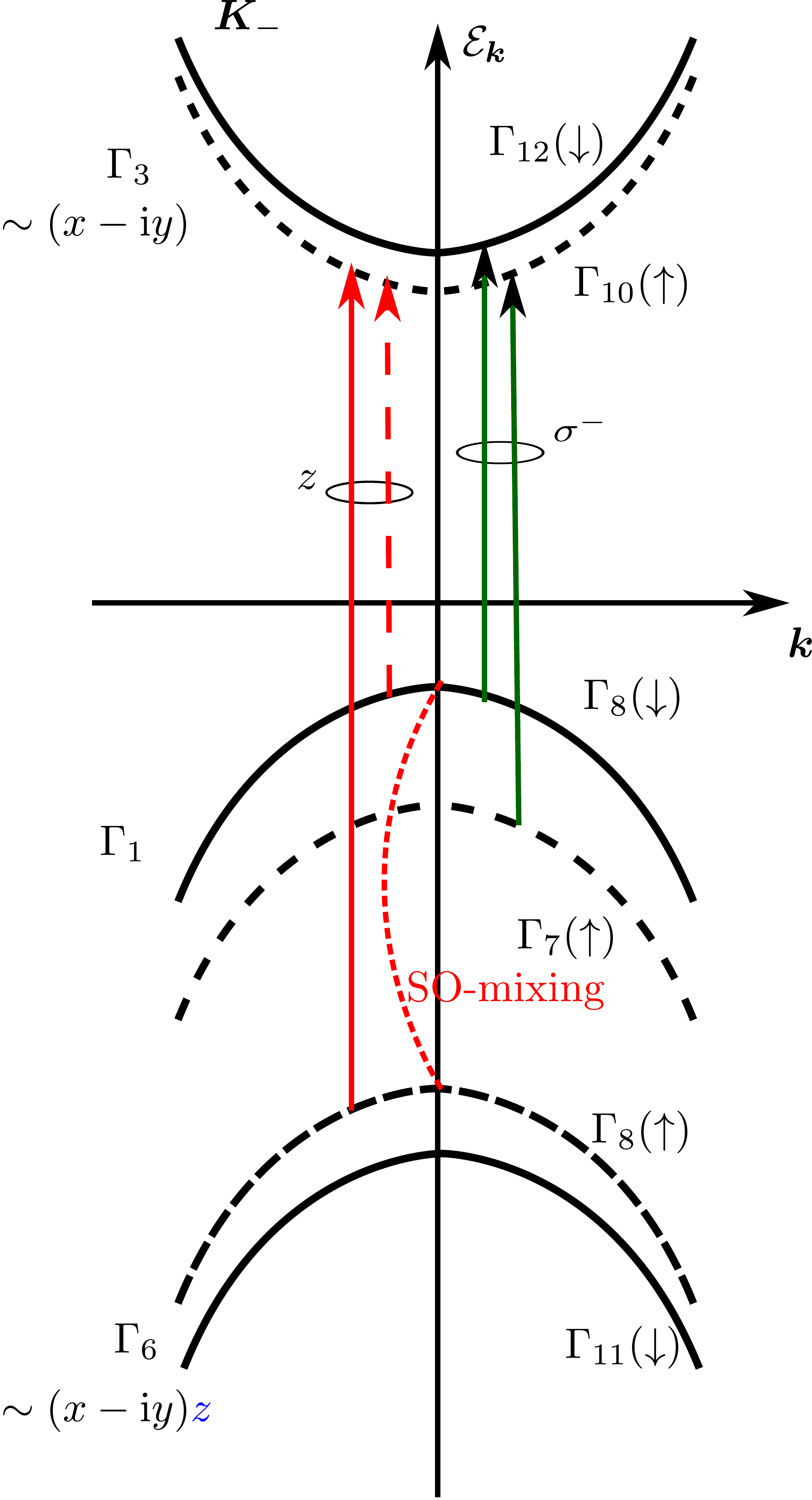}
\caption{Schematics of the bandstructure at the $\bm K_-$ with the remote valence bands shown. Irreducible representations are those of $C_{3h}$ point symmetry group. The red and green arrows denote the optical transitions allowed in $z$ and in-plane polarizations, respectively. The spin-orbit mixing between the valence bands is shown by the red dotted line.}\label{fig:S0} 
\end{figure}

The excitons formed in the course of optical transitions involving topmost valence band and the top conduction subband, e.g., vertical spin-conserving transitions, Fig.~1(e,f) of the main text and Fig.~\ref{fig:S0}, transform according to the reducible representation
\begin{equation}
\label{spin:cons}
\Gamma_{X,\parallel}^{D_{3h}} = \Gamma_7^{D_{3h}} \times \Gamma_9^{D_{3h}} = \Gamma_5^{D_{3h}} + \Gamma_6^{D_{3h}}.
\end{equation}
The in-plane components of the vector $(x,y)$ transform according to the irreducible representation $\Gamma_6^{D_{3h}}$ resulting in the optical activity of the excitons denoted by green arrows in Fig.~1(e,f) of the main text. Two remaining optical transitions where the electron from the top conduction subband recombines with the unoccupied state in the other valley are forbidden.

Let us now consider the excitons formed in the course of optical transitions involving the bottom conduction subband, e.g, vertical transitions with antiparallel conduction and valence band states, Fig.~1(f) of the main text and Fig.~\ref{fig:S0}. We have
\begin{equation}
\label{spin:cons}
\Gamma_{X,\uparrow\downarrow}^{D_{3h}} = \Gamma_7^{D_{3h}} \times \Gamma_8^{D_{3h}} = \Gamma_3^{D_{3h}}+ \Gamma_4^{D_{3h}} + \Gamma_6^{D_{3h}}.
\end{equation}
The $z$-component of a vector transforms according to the $\Gamma_4^{D_{3h}}$ irreducible representation.
The states transforming according to the $\Gamma_6^{D_{3h}}$ correspond to intervalley transitions activated with symmetric ($\Gamma_1^{D_{3h}}$) phonon or due to the localization of the exciton on axially-symmetric defect potential~\cite{Wang:2015d}. We are interested in two remaining states transforming according to the irreducible representations $\Gamma_3^{D_{3h}}$ and  $\Gamma_4^{D_{3h}}$, respectively. These states are the linear combinations of the excitons formed in the course of vertical transitions between the states with opposite spins in the $\bm K_+$ and $\bm K_-$ valley. One of the states ($\Gamma_3^{D_{3h}}$) is forbidden and another one ($\Gamma_4^{D_{3h}}$) is allowed in $z$-polarization as depicted in Fig.~1(f) of the main text by red arrows.

\subsection{Spin-orbit mixing of bands}

The origin of the interband transitions without electron spin conservation is the spin-orbit interaction induced band mixing~\cite{Pikus1988,Kormanyos:2015a}. To illustrate the effect we consider the mixing of the topmost valence band $v$ and the remote valence band $v'$ ($vb-1$ in the notations of Ref.~\cite{Kormanyos:2015a}) whose orbital Bloch functions transform at the $\bm K_\pm$ points as
\begin{subequations}
\label{E''}
\begin{align}
&\bm K_+: &\Gamma_5^{C_{3h}} = \Gamma_2^{C_{3h}} \times \Gamma_4^{C_{3h}},~~&\mathcal U_{v'}^+ \sim z(x+\mathrm i y) \label{1}\\
&\bm K_-: &\Gamma_6^{C_{3h}} = \Gamma_3^{C_{3h}} \times \Gamma_4^{C_{3h}},~~&\mathcal U_{v'}^- \sim z(x-\mathrm i y) \label{2}.
\end{align}
\end{subequations}
Here and in what follows we use the irreducible representations of $C_{3h}$ point group and omit the superscript denoting the group. The orbital Bloch functions $\mathcal U_{v'}^\pm$ are odd at the horizontal reflection unlike the even Bloch states of $cb+2$ and $vb-3$ involved in the exciton mixing~\cite{PhysRevB.95.035311}, while at the operations which do not involve $z\to-z$ these functions transform like the conduction band Bloch functions. With account for spin the remote $v'$ valence band gives rise to the following spin subbands:
\begin{subequations}
\label{products1}
\begin{align}
\Gamma_5 \times \Gamma_7 = \Gamma_{12} \quad  \mathcal U^+_{v'(\uparrow)} \sim (x+\mathrm i y) z\uparrow,\\
\Gamma_6 \times \Gamma_7 = \Gamma_{8} \quad  \mathcal U^-_{v'(\uparrow)}\sim (x-\mathrm i y) z\uparrow,\\
\Gamma_5 \times \Gamma_8 = \Gamma_{7} \quad  \mathcal U^+_{v'(\downarrow)}\sim (x+\mathrm i y) z\downarrow,\\
\Gamma_6 \times \Gamma_8 = \Gamma_{11} \quad  \mathcal U^-_{v'(\uparrow)}\sim (x-\mathrm i y) z\downarrow.
\end{align}
\end{subequations}
Hence, in each valley there are deep valence band states which transform according to the same representations as the valence band top, but with reversed spins, see Fig.~\ref{fig:S0} for illustration of the states in the $\bm K_-$ valley.

The Bloch functions of the top valence band, $\mathcal U^-_{\Gamma_8(\downarrow)}$ [$\mathcal U^+_{\Gamma_7(\uparrow)}$], and the remote band, $\mathcal U^-_{v'(\uparrow)}$ [$\mathcal U_{v'(\uparrow)}^+$], at the $\bm K_-$ [$\bm K_+$] valley transform according to the same irreducible representation of the $C_{3h}$ point group, therefore, these states are mixed, in general. The origin of the mixing is the atomic spin-orbit coupling in the form of $$\bm L \cdot \bm S,$$ where $\bm L$ and $\bm S$ are the orbital and spin operators. For example, the state $\mathcal U^-_{v'(\uparrow)}$ transforms as the state with the total angular momentum $z$-component, $F_z = L_z + S_z = -1+1/2=1/2$, which is exactly the same as the $z$-component of the topmost valence band state in the same valley, $\mathcal U^-_{\Gamma_8(\downarrow)}$ transforming as the function with $F_z = 0+1/2 = 1/2$.
 Within the effective Hamiltonian method the valence band top wavefunction in $\bm K_\pm$ valleys can be presented as
\begin{subequations}
\label{top}
\begin{equation}
\label{top+}
\alpha \mathcal U_{\Gamma_7(\uparrow)}^{+} + \beta \mathcal U^+_{v'(\downarrow)} \quad (\bm K_+\mbox{-valley}),
\end{equation}
\begin{equation}
\label{top-}
\alpha \mathcal U_{\Gamma_8(\uparrow)}^{-} + \beta \mathcal U^-_{v'(\downarrow)} \quad (\bm K_-\mbox{-valley}),
\end{equation}
\end{subequations}
In the first order perturbation theory one has
\begin{equation}
\label{beta}
\alpha = 1, \quad \beta = \frac{\Delta_{v,v-1}}{E_v - E_{v-1}},
\end{equation}

where $\Delta_{v,v-1}$ is the spin-orbit mixing matrix elements and $E_i$ are the corresponding band edges.
Due to the admixture, Eqs.~\eqref{top}, \eqref{beta}, the transition in $z$ polarization becomes possible between the valence and conduction band states with opposite spin orientations, see red dashed arrow in Fig.~\ref{fig:S0}: This is because the transition between the admixed $\mathcal U_{v'}^\pm$ state and the conduction band state is possible in $z$ polarization and the spins of the admixed state and the conduction band state are parallel. The ratio of the oscillator strengths (radiative dampings) of the ``dark'' exciton in the $z$ polarization and of the ``bright'' exciton in the in-plane $x$ polarization can be estimated as
\begin{equation}
\label{radiative}
\gamma = \frac{\Gamma_{0,\rm dark}}{\Gamma_{0,\rm bright}} = \frac{|d_\perp|^2}{|d_\parallel|^2} \beta^2,
\end{equation}
where $d_\perp$ is the interband ($v'\to c$) electric dipole matrix element in the $z$ polarization and $d_\parallel$ is the interband ($v\to c$) electric dipole moment matrix element in the in-plane polarization and we neglected the difference of effective masses in the different spin subbands which may affect the exciton envelope functions. Assuming $|d_\perp|^2/|d_\parallel|^2 \sim 1$ and putting $E_v - E_{v-1}\sim 1$~eV we obtain, for $\Delta_{v,v-1}$ in the range of $10\ldots 100$~meV, the damping ratio  in the range of $\gamma=10^{-4}\ldots 10^{-2}$. The density functional theory calculations gives $\gamma\sim 10^{-3}$ in reasonable agreement with the crude estimate above~\cite{Echeverry:2016}. We note that additive contribution to the mixing and the optical transitions is provided by the spin-orbit coupling with remote conduction band states. The parameters of these remote bands are not well established therefore the estimate Eq.~\eqref{radiative} can serve as a criterion to improve the parameterizations of the effective Hamiltonian models.

\newpage
\subsection{Additional experimental data}
\begin{figure}[htbp]
\includegraphics [width=0.8\linewidth,keepaspectratio=true]{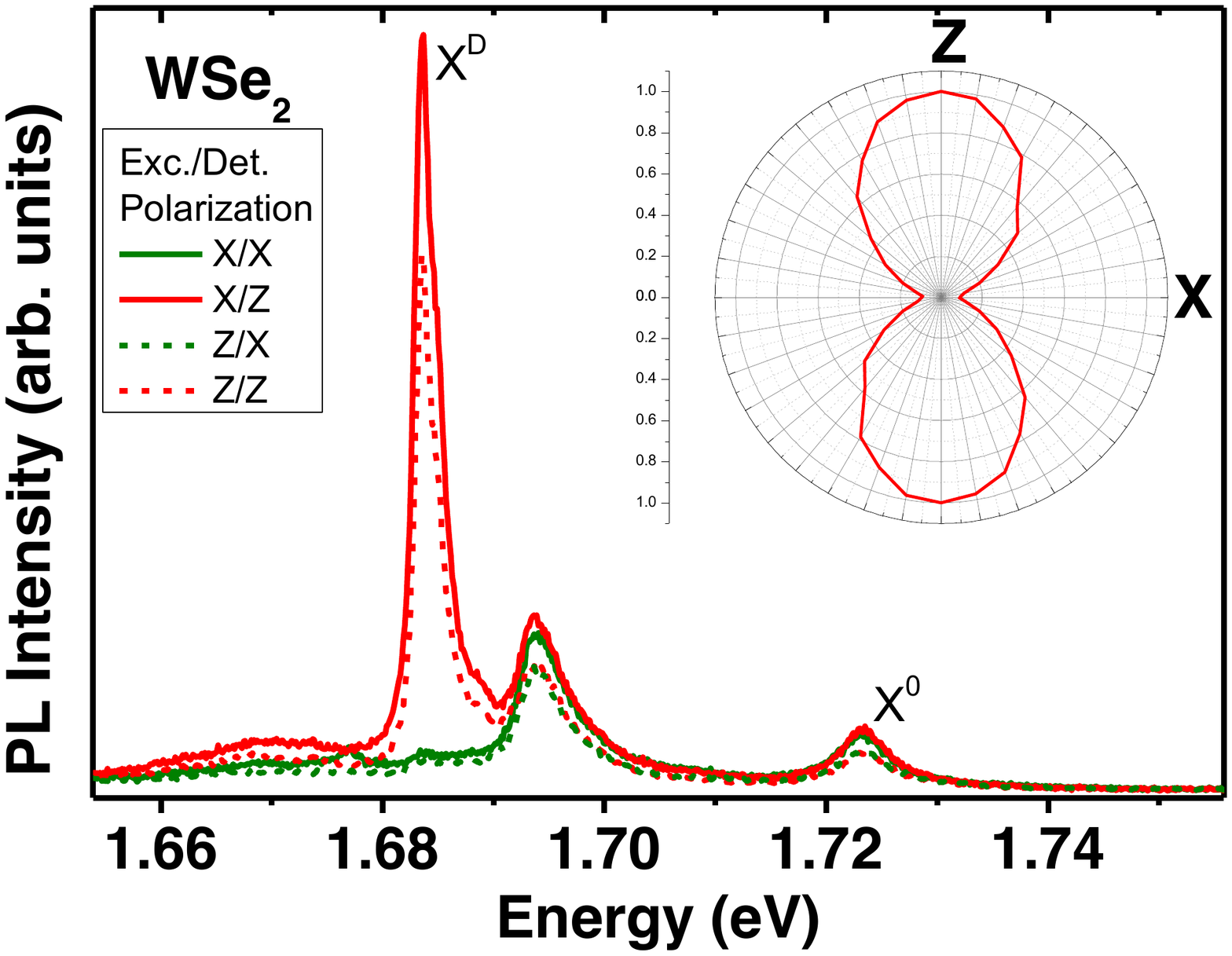}
\caption{Photoluminescence spectra of WSe$_2$ monolayer in the edge excitation/detection geometry for different configurations of excitation and detection polarizations. The inset shows the polar plot of the intensity of the X$^D$ transition for $x$-polarized excitation and varying detection polarization}\label{fig:S1} 
\end{figure}

Figure~\ref{fig:S1} displays the photoluminescence spectra of WSe$_2$ in different plarization configurations together with a polar plot of intensity of X$^D$ transition for fixed excitation and varied detection polarization.\\

\begin{figure}[htbp]
\includegraphics [width=0.8\linewidth,keepaspectratio=true]{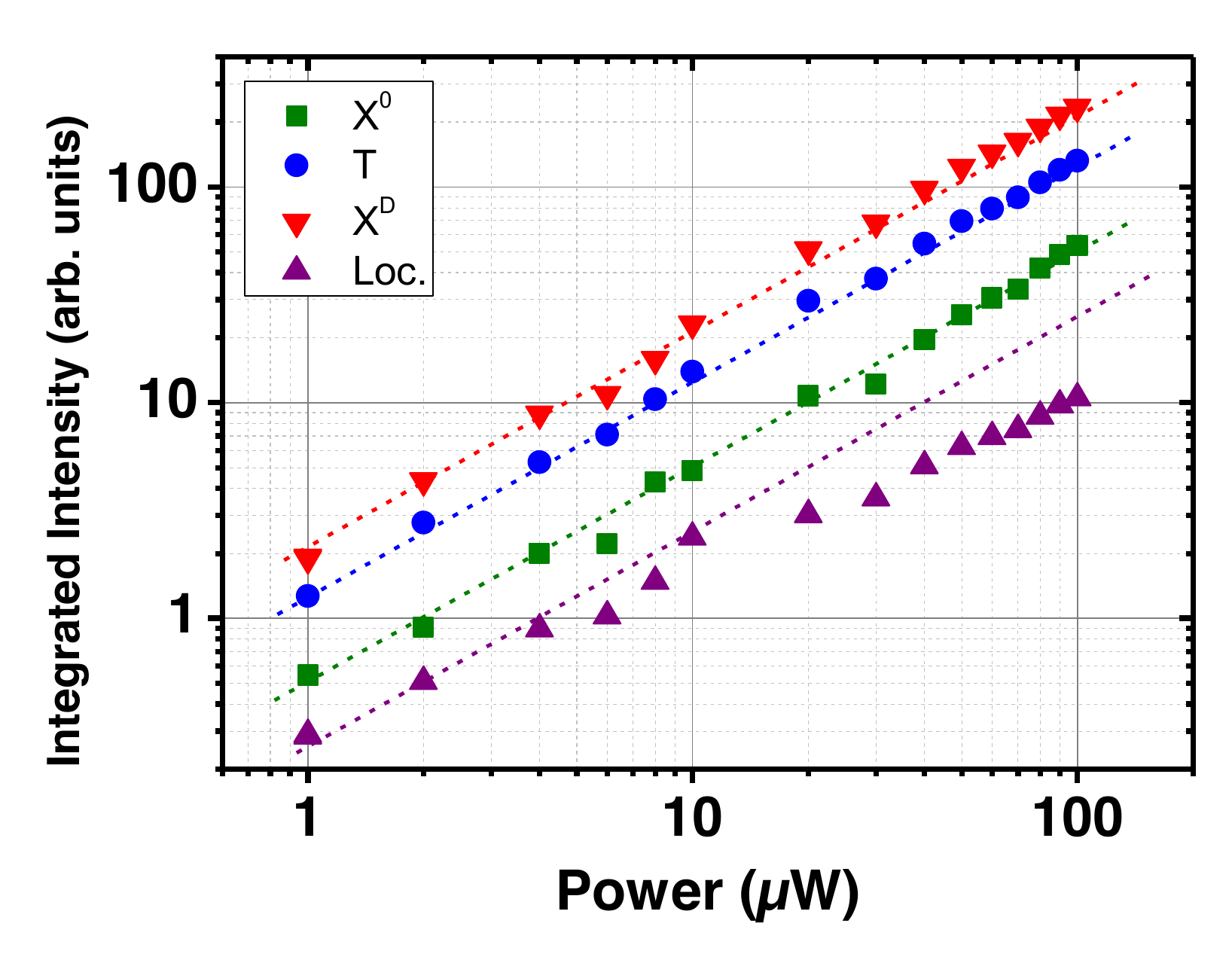}
\caption{Intensities of X$^0$, T, X$^D$ and localized (Loc.) excitons transitions as a function of power for a WSe$_2$ monolayer in the edge excitation/detection geometry. Excitation (detection) polarizations are $x$ ($z$). Localized excitons intensity is obtained by integrating the spectra for energies below 1.68 eV. The dashed lines show linear variation. X$^0$, T, and X$^D$ scale linearly with power while localized excitons saturate. This indicates that the X$^D$ transition is not related to a state localized on a defect.}\label{fig:S2} 
\end{figure}

Figure~\ref{fig:S2} presents the intensities of emission of neutral (X$^0$), charged (T), "Dark" (X$^D$) and localized (Loc.) excitons as a function of excitation power.\\

\begin{figure}[htbp]
\includegraphics [width=0.5\textwidth,keepaspectratio=true]{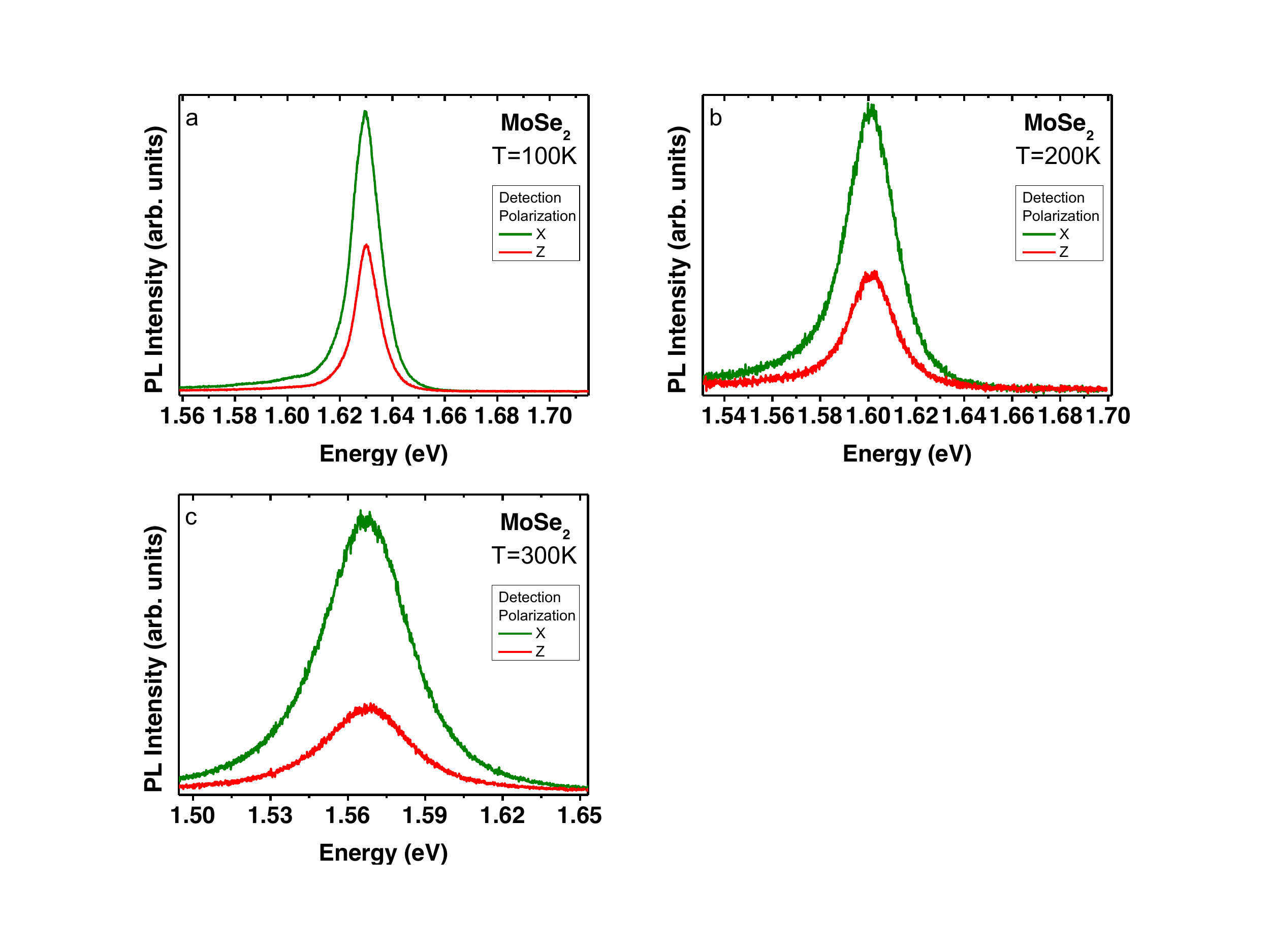}
\caption{Photoluminescence spectra of MoSe$_2$ monolayer in the edge excitation/detection geometry for $x$ and $z$ polarized detection at (a) 100 K, (b) 200 K, (c) 300 K. We do not detect any fingerprint of the X$^D$ transition.}\label{fig:S3} 
\end{figure}

Figure~\ref{fig:S3} shows the photoluminescence spectrum of the MoSe$_2$ monolayer at different temperatures.\\

\subsection{Spatial analysis of the PL}
\label{sec:spatial analysis of the PL}
\begin{figure*}
\includegraphics [width=0.9\textwidth,keepaspectratio=true]{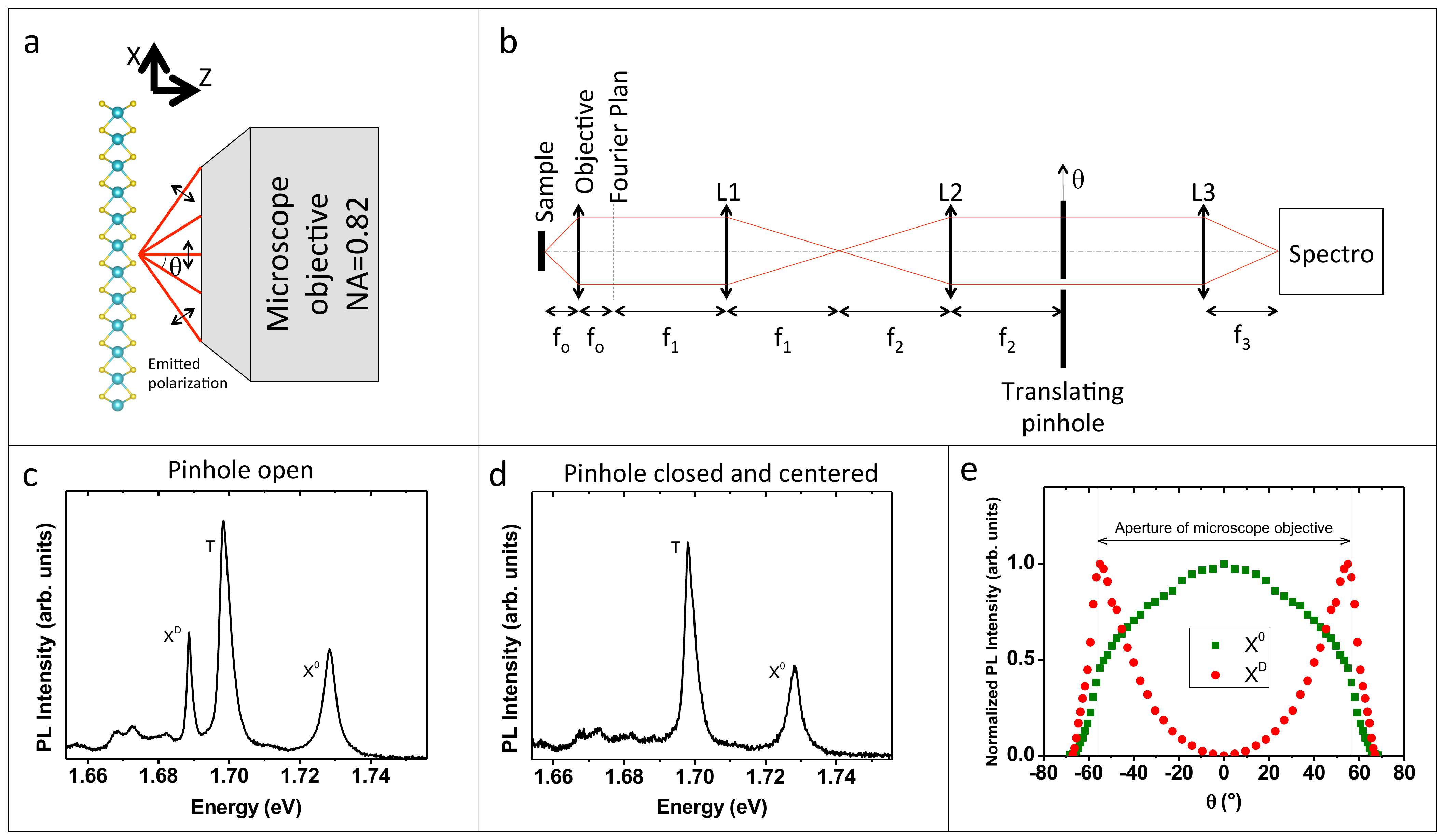}
\caption{Light propagating perpendicular to the monolayer plane - detection of the photoluminescence from the top of the sample. Spatial analysis of the photoluminescence spot.
(a) Sketch showing the surface detection configuration. Because of the high numerical aperture, light with z polarization can be detected at large angles $\theta$ . (b) Sketch of the optical setup enabling filtering of detected angles $\theta$. (c) Photoluminescence spectrum of WS$_2$ monolayer when the pinhole is opened (no angle filtering). (d) Same as (c) but for the pinhole closed at its maximum and centered (only light with $\theta{=0}$ is detected). In this configuration the X$^D$ transition vanishes. (e) Intensities of X$^0$ and X$^D$ transitions as a function of the filtered angle. X$^D$ transition is only visible for large $\theta$.}\label{fig:S4} 
\end{figure*}

In Fig 1c of the main text, we showed that the X$^D$ transition at 1.68 eV is visible even for excitation and detection from the surface of the WS$_2$ ML. This may be surprising at the first sight because the light propagation vector is along the {$z$-axis} in this configuration so that the X$^D$ transition (which is $z$-polarized) should not be visible. We show here that this is due to our large numerical aperture microscope objective. Fig.~\ref{fig:S4}(a) sketches the detection configuration: Because of the large numerical aperture, light with an in-plane wave-vector component can be detected on the edge of the objective aperture. To confirm this situation we selectively filter the emitted light in angle by imaging the Fourier plan of the microscope objective [lenses L1 and L2 of Fig.~\ref{fig:S4}(b)] and by placing a pinhole on this image. Translating the pinhole selects the angle of detection. Figure~\ref{fig:S4}(c) shows the PL spectrum when the pinhole is opened, the configuration is thus the same as in Fig.~1(c) of the main text where both X$^0$ and X$^D$ transitions are visible. When the pinhole is centered and almost closed, Fig.~\ref{fig:S4}(d), only light with out-of-plane wavevector is detected. Consequently the X$^D$ transition is absent. In Fig.~\ref{fig:S4}(e), we plot the intensities of both X$^0$ and X$^D$ transitions as a function of the emitted angle $\theta$. It clearly show that X$^D$ transition is only visible for large $\theta$ whereas the X$^0$ is maximum for $\theta=0$.

\begin{thebibliography}{51}%
\makeatletter
\providecommand \@ifxundefined [1]{%
 \@ifx{#1\undefined}
}%
\providecommand \@ifnum [1]{%
 \ifnum #1\expandafter \@firstoftwo
 \else \expandafter \@secondoftwo
 \fi
}%
\providecommand \@ifx [1]{%
 \ifx #1\expandafter \@firstoftwo
 \else \expandafter \@secondoftwo
 \fi
}%
\providecommand \natexlab [1]{#1}%
\providecommand \enquote  [1]{``#1''}%
\providecommand \bibnamefont  [1]{#1}%
\providecommand \bibfnamefont [1]{#1}%
\providecommand \citenamefont [1]{#1}%
\providecommand \href@noop [0]{\@secondoftwo}%
\providecommand \href [0]{\begingroup \@sanitize@url \@href}%
\providecommand \@href[1]{\@@startlink{#1}\@@href}%
\providecommand \@@href[1]{\endgroup#1\@@endlink}%
\providecommand \@sanitize@url [0]{\catcode `\\12\catcode `\$12\catcode
  `\&12\catcode `\#12\catcode `\^12\catcode `\_12\catcode `\%12\relax}%
\providecommand \@@startlink[1]{}%
\providecommand \@@endlink[0]{}%
\providecommand \url  [0]{\begingroup\@sanitize@url \@url }%
\providecommand \@url [1]{\endgroup\@href {#1}{\urlprefix }}%
\providecommand \urlprefix  [0]{URL }%
\providecommand \Eprint [0]{\href }%
\providecommand \doibase [0]{http://dx.doi.org/}%
\providecommand \selectlanguage [0]{\@gobble}%
\providecommand \bibinfo  [0]{\@secondoftwo}%
\providecommand \bibfield  [0]{\@secondoftwo}%
\providecommand \translation [1]{[#1]}%
\providecommand \BibitemOpen [0]{}%
\providecommand \bibitemStop [0]{}%
\providecommand \bibitemNoStop [0]{.\EOS\space}%
\providecommand \EOS [0]{\spacefactor3000\relax}%
\providecommand \BibitemShut  [1]{\csname bibitem#1\endcsname}%
\let\auto@bib@innerbib\@empty
\bibitem [{\citenamefont {Butler}\ \emph {et~al.}(2013)\citenamefont {Butler},
  \citenamefont {Hollen}, \citenamefont {Cao}, \citenamefont {Cui},
  \citenamefont {Gupta}, \citenamefont {Guti{\'e}rrez}, \citenamefont {Heinz},
  \citenamefont {Hong}, \citenamefont {Huang}, \citenamefont {Ismach},
  \citenamefont {Johnston-Halperin}, \citenamefont {Kuno}, \citenamefont
  {Plashnitsa}, \citenamefont {Robinson}, \citenamefont {Ruoff}, \citenamefont
  {Salahuddin}, \citenamefont {Shan}, \citenamefont {Shi}, \citenamefont
  {Spencer}, \citenamefont {Terrones}, \citenamefont {Windl},\ and\
  \citenamefont {Goldberger}}]{Butler:2013a}%
  \BibitemOpen
  \bibfield  {author} {\bibinfo {author} {\bibfnamefont {Sheneve~Z.}\
  \bibnamefont {Butler}}, \bibinfo {author} {\bibfnamefont {Shawna~M.}\
  \bibnamefont {Hollen}}, \bibinfo {author} {\bibfnamefont {Linyou}\
  \bibnamefont {Cao}}, \bibinfo {author} {\bibfnamefont {Yi}~\bibnamefont
  {Cui}}, \bibinfo {author} {\bibfnamefont {Jay~A.}\ \bibnamefont {Gupta}},
  \bibinfo {author} {\bibfnamefont {Humberto~R.}\ \bibnamefont
  {Guti{\'e}rrez}}, \bibinfo {author} {\bibfnamefont {Tony~F.}\ \bibnamefont
  {Heinz}}, \bibinfo {author} {\bibfnamefont {Seung~Sae}\ \bibnamefont {Hong}},
  \bibinfo {author} {\bibfnamefont {Jiaxing}\ \bibnamefont {Huang}}, \bibinfo
  {author} {\bibfnamefont {Ariel~F.}\ \bibnamefont {Ismach}}, \bibinfo {author}
  {\bibfnamefont {Ezekiel}\ \bibnamefont {Johnston-Halperin}}, \bibinfo
  {author} {\bibfnamefont {Masaru}\ \bibnamefont {Kuno}}, \bibinfo {author}
  {\bibfnamefont {Vladimir~V.}\ \bibnamefont {Plashnitsa}}, \bibinfo {author}
  {\bibfnamefont {Richard~D.}\ \bibnamefont {Robinson}}, \bibinfo {author}
  {\bibfnamefont {Rodney~S.}\ \bibnamefont {Ruoff}}, \bibinfo {author}
  {\bibfnamefont {Sayeef}\ \bibnamefont {Salahuddin}}, \bibinfo {author}
  {\bibfnamefont {Jie}\ \bibnamefont {Shan}}, \bibinfo {author} {\bibfnamefont
  {Li}~\bibnamefont {Shi}}, \bibinfo {author} {\bibfnamefont {Michael~G.}\
  \bibnamefont {Spencer}}, \bibinfo {author} {\bibfnamefont {Mauricio}\
  \bibnamefont {Terrones}}, \bibinfo {author} {\bibfnamefont {Wolfgang}\
  \bibnamefont {Windl}}, \ and\ \bibinfo {author} {\bibfnamefont {Joshua~E.}\
  \bibnamefont {Goldberger}},\ }\bibfield  {title} {\enquote {\bibinfo {title}
  {Progress, challenges, and opportunities in two-dimensional materials beyond
  graphene},}\ }\href@noop {} {\bibfield  {journal} {\bibinfo  {journal} {ACS
  Nano}\ }\textbf {\bibinfo {volume} {7}},\ \bibinfo {pages} {2898--2926}
  (\bibinfo {year} {2013})}\BibitemShut {NoStop}%
\bibitem [{\citenamefont {Geim}\ and\ \citenamefont
  {Grigorieva}(2013)}]{Geim:2013a}%
  \BibitemOpen
  \bibfield  {author} {\bibinfo {author} {\bibfnamefont {A.~K.}\ \bibnamefont
  {Geim}}\ and\ \bibinfo {author} {\bibfnamefont {I.~V.}\ \bibnamefont
  {Grigorieva}},\ }\bibfield  {title} {\enquote {\bibinfo {title} {Van der
  waals heterostructures},}\ }\href {\doibase doi:10.1038/nature12385}
  {\bibfield  {journal} {\bibinfo  {journal} {Nature}\ }\textbf {\bibinfo
  {volume} {499}},\ \bibinfo {pages} {419---425} (\bibinfo {year}
  {2013})}\BibitemShut {NoStop}%
\bibitem [{\citenamefont {Mak}\ \emph {et~al.}(2010)\citenamefont {Mak},
  \citenamefont {Lee}, \citenamefont {Hone}, \citenamefont {Shan},\ and\
  \citenamefont {Heinz}}]{Mak:2010a}%
  \BibitemOpen
  \bibfield  {author} {\bibinfo {author} {\bibfnamefont {Kin~Fai}\ \bibnamefont
  {Mak}}, \bibinfo {author} {\bibfnamefont {Changgu}\ \bibnamefont {Lee}},
  \bibinfo {author} {\bibfnamefont {James}\ \bibnamefont {Hone}}, \bibinfo
  {author} {\bibfnamefont {Jie}\ \bibnamefont {Shan}}, \ and\ \bibinfo {author}
  {\bibfnamefont {Tony~F.}\ \bibnamefont {Heinz}},\ }\bibfield  {title}
  {\enquote {\bibinfo {title} {Atomically thin ${\mathrm{mos}}_{2}$: A new
  direct-gap semiconductor},}\ }\href@noop {} {\bibfield  {journal} {\bibinfo
  {journal} {Phys. Rev. Lett.}\ }\textbf {\bibinfo {volume} {105}},\ \bibinfo
  {pages} {136805} (\bibinfo {year} {2010})}\BibitemShut {NoStop}%
\bibitem [{\citenamefont {Splendiani}\ \emph {et~al.}(2010)\citenamefont
  {Splendiani}, \citenamefont {Sun}, \citenamefont {Zhang}, \citenamefont {Li},
  \citenamefont {Kim}, \citenamefont {Chim}, \citenamefont {Galli},\ and\
  \citenamefont {Wang}}]{Splendiani:2010a}%
  \BibitemOpen
  \bibfield  {author} {\bibinfo {author} {\bibfnamefont {Andrea}\ \bibnamefont
  {Splendiani}}, \bibinfo {author} {\bibfnamefont {Liang}\ \bibnamefont {Sun}},
  \bibinfo {author} {\bibfnamefont {Yuanbo}\ \bibnamefont {Zhang}}, \bibinfo
  {author} {\bibfnamefont {Tianshu}\ \bibnamefont {Li}}, \bibinfo {author}
  {\bibfnamefont {Jonghwan}\ \bibnamefont {Kim}}, \bibinfo {author}
  {\bibfnamefont {Chi-Yung}\ \bibnamefont {Chim}}, \bibinfo {author}
  {\bibfnamefont {Giulia}\ \bibnamefont {Galli}}, \ and\ \bibinfo {author}
  {\bibfnamefont {Feng}\ \bibnamefont {Wang}},\ }\bibfield  {title} {\enquote
  {\bibinfo {title} {Emerging photoluminescence in monolayer mos2},}\
  }\href@noop {} {\bibfield  {journal} {\bibinfo  {journal} {Nano Letters}\
  }\textbf {\bibinfo {volume} {10}},\ \bibinfo {pages} {1271} (\bibinfo {year}
  {2010})}\BibitemShut {NoStop}%
\bibitem [{\citenamefont {Wang}\ \emph {et~al.}(2012)\citenamefont {Wang},
  \citenamefont {Kalantar-Zadeh}, \citenamefont {Kis}, \citenamefont
  {Coleman},\ and\ \citenamefont {Strano}}]{Wang:2012c}%
  \BibitemOpen
  \bibfield  {author} {\bibinfo {author} {\bibfnamefont {Qing~Hua}\
  \bibnamefont {Wang}}, \bibinfo {author} {\bibfnamefont {Kourosh}\
  \bibnamefont {Kalantar-Zadeh}}, \bibinfo {author} {\bibfnamefont {Andras}\
  \bibnamefont {Kis}}, \bibinfo {author} {\bibfnamefont {Jonathan~N}\
  \bibnamefont {Coleman}}, \ and\ \bibinfo {author} {\bibfnamefont {Michael~S}\
  \bibnamefont {Strano}},\ }\bibfield  {title} {\enquote {\bibinfo {title}
  {Electronics and optoelectronics of two-dimensional transition metal
  dichalcogenides},}\ }\href@noop {} {\bibfield  {journal} {\bibinfo  {journal}
  {Nature nanotechnology}\ }\textbf {\bibinfo {volume} {7}},\ \bibinfo {pages}
  {699--712} (\bibinfo {year} {2012})}\BibitemShut {NoStop}%
\bibitem [{\citenamefont {He}\ \emph {et~al.}(2014)\citenamefont {He},
  \citenamefont {Kumar}, \citenamefont {Zhao}, \citenamefont {Wang},
  \citenamefont {Mak}, \citenamefont {Zhao},\ and\ \citenamefont
  {Shan}}]{He:2014a}%
  \BibitemOpen
  \bibfield  {author} {\bibinfo {author} {\bibfnamefont {Keliang}\ \bibnamefont
  {He}}, \bibinfo {author} {\bibfnamefont {Nardeep}\ \bibnamefont {Kumar}},
  \bibinfo {author} {\bibfnamefont {Liang}\ \bibnamefont {Zhao}}, \bibinfo
  {author} {\bibfnamefont {Zefang}\ \bibnamefont {Wang}}, \bibinfo {author}
  {\bibfnamefont {Kin~Fai}\ \bibnamefont {Mak}}, \bibinfo {author}
  {\bibfnamefont {Hui}\ \bibnamefont {Zhao}}, \ and\ \bibinfo {author}
  {\bibfnamefont {Jie}\ \bibnamefont {Shan}},\ }\bibfield  {title} {\enquote
  {\bibinfo {title} {Tightly bound excitons in monolayer
  ${\mathrm{wse}}_{2}$},}\ }\href {\doibase 10.1103/PhysRevLett.113.026803}
  {\bibfield  {journal} {\bibinfo  {journal} {Phys. Rev. Lett.}\ }\textbf
  {\bibinfo {volume} {113}},\ \bibinfo {pages} {026803} (\bibinfo {year}
  {2014})}\BibitemShut {NoStop}%
\bibitem [{\citenamefont {{Ugeda}}\ \emph {et~al.}(2014)\citenamefont
  {{Ugeda}}, \citenamefont {{Bradley}}, \citenamefont {{Shi}}, \citenamefont
  {{da Jornada}}, \citenamefont {{Zhang}}, \citenamefont {{Qiu}}, \citenamefont
  {{Mo}}, \citenamefont {{Hussain}}, \citenamefont {{Shen}}, \citenamefont
  {{Wang}}, \citenamefont {{Louie}},\ and\ \citenamefont
  {{Crommie}}}]{Ugeda:2014a}%
  \BibitemOpen
  \bibfield  {author} {\bibinfo {author} {\bibfnamefont {M.~M.}\ \bibnamefont
  {{Ugeda}}}, \bibinfo {author} {\bibfnamefont {A.~J.}\ \bibnamefont
  {{Bradley}}}, \bibinfo {author} {\bibfnamefont {S.-F.}\ \bibnamefont
  {{Shi}}}, \bibinfo {author} {\bibfnamefont {F.~H.}\ \bibnamefont {{da
  Jornada}}}, \bibinfo {author} {\bibfnamefont {Y.}~\bibnamefont {{Zhang}}},
  \bibinfo {author} {\bibfnamefont {D.~Y.}\ \bibnamefont {{Qiu}}}, \bibinfo
  {author} {\bibfnamefont {S.-K.}\ \bibnamefont {{Mo}}}, \bibinfo {author}
  {\bibfnamefont {Z.}~\bibnamefont {{Hussain}}}, \bibinfo {author}
  {\bibfnamefont {Z.-X.}\ \bibnamefont {{Shen}}}, \bibinfo {author}
  {\bibfnamefont {F.}~\bibnamefont {{Wang}}}, \bibinfo {author} {\bibfnamefont
  {S.~G.}\ \bibnamefont {{Louie}}}, \ and\ \bibinfo {author} {\bibfnamefont
  {M.~F.}\ \bibnamefont {{Crommie}}},\ }\bibfield  {title} {\enquote {\bibinfo
  {title} {Observation of giant bandgap renormalization and excitonic effects
  in a monolayer transition metal dichalcogenide semiconductor},}\ }\href@noop
  {} {\bibfield  {journal} {\bibinfo  {journal} {Nature Materials}\ }\textbf
  {\bibinfo {volume} {13}},\ \bibinfo {pages} {1091--1095} (\bibinfo {year}
  {2014})}\BibitemShut {NoStop}%
\bibitem [{\citenamefont {Chernikov}\ \emph {et~al.}(2014)\citenamefont
  {Chernikov}, \citenamefont {Berkelbach}, \citenamefont {Hill}, \citenamefont
  {Rigosi}, \citenamefont {Li}, \citenamefont {Aslan}, \citenamefont
  {Reichman}, \citenamefont {Hybertsen},\ and\ \citenamefont
  {Heinz}}]{Chernikov:2014a}%
  \BibitemOpen
  \bibfield  {author} {\bibinfo {author} {\bibfnamefont {Alexey}\ \bibnamefont
  {Chernikov}}, \bibinfo {author} {\bibfnamefont {Timothy~C.}\ \bibnamefont
  {Berkelbach}}, \bibinfo {author} {\bibfnamefont {Heather~M.}\ \bibnamefont
  {Hill}}, \bibinfo {author} {\bibfnamefont {Albert}\ \bibnamefont {Rigosi}},
  \bibinfo {author} {\bibfnamefont {Yilei}\ \bibnamefont {Li}}, \bibinfo
  {author} {\bibfnamefont {Ozgur~Burak}\ \bibnamefont {Aslan}}, \bibinfo
  {author} {\bibfnamefont {David~R.}\ \bibnamefont {Reichman}}, \bibinfo
  {author} {\bibfnamefont {Mark~S.}\ \bibnamefont {Hybertsen}}, \ and\ \bibinfo
  {author} {\bibfnamefont {Tony~F.}\ \bibnamefont {Heinz}},\ }\bibfield
  {title} {\enquote {\bibinfo {title} {Exciton binding energy and nonhydrogenic
  rydberg series in monolayer ${\mathrm{ws}}_{2}$},}\ }\href {\doibase
  10.1103/PhysRevLett.113.076802} {\bibfield  {journal} {\bibinfo  {journal}
  {Phys. Rev. Lett.}\ }\textbf {\bibinfo {volume} {113}},\ \bibinfo {pages}
  {076802} (\bibinfo {year} {2014})}\BibitemShut {NoStop}%
\bibitem [{\citenamefont {{Ye}}\ \emph {et~al.}(2014)\citenamefont {{Ye}},
  \citenamefont {{Cao}}, \citenamefont {{O'Brien}}, \citenamefont {{Zhu}},
  \citenamefont {{Yin}}, \citenamefont {{Wang}}, \citenamefont {{Louie}},\ and\
  \citenamefont {{Zhang}}}]{Ye:2014a}%
  \BibitemOpen
  \bibfield  {author} {\bibinfo {author} {\bibfnamefont {Z.}~\bibnamefont
  {{Ye}}}, \bibinfo {author} {\bibfnamefont {T.}~\bibnamefont {{Cao}}},
  \bibinfo {author} {\bibfnamefont {K.}~\bibnamefont {{O'Brien}}}, \bibinfo
  {author} {\bibfnamefont {H.}~\bibnamefont {{Zhu}}}, \bibinfo {author}
  {\bibfnamefont {X.}~\bibnamefont {{Yin}}}, \bibinfo {author} {\bibfnamefont
  {Y.}~\bibnamefont {{Wang}}}, \bibinfo {author} {\bibfnamefont {S.~G.}\
  \bibnamefont {{Louie}}}, \ and\ \bibinfo {author} {\bibfnamefont
  {X.}~\bibnamefont {{Zhang}}},\ }\bibfield  {title} {\enquote {\bibinfo
  {title} {{Probing Excitonic Dark States in Single-layer Tungsten
  Disulfide}},}\ }\href@noop {} {\bibfield  {journal} {\bibinfo  {journal}
  {Nature}\ }\textbf {\bibinfo {volume} {513}},\ \bibinfo {pages} {214--218}
  (\bibinfo {year} {2014})}\BibitemShut {NoStop}%
\bibitem [{\citenamefont {Qiu}\ \emph {et~al.}(2013)\citenamefont {Qiu},
  \citenamefont {da~Jornada},\ and\ \citenamefont {Louie}}]{Qiu:2013a}%
  \BibitemOpen
  \bibfield  {author} {\bibinfo {author} {\bibfnamefont {Diana~Y.}\
  \bibnamefont {Qiu}}, \bibinfo {author} {\bibfnamefont {Felipe~H.}\
  \bibnamefont {da~Jornada}}, \ and\ \bibinfo {author} {\bibfnamefont
  {Steven~G.}\ \bibnamefont {Louie}},\ }\bibfield  {title} {\enquote {\bibinfo
  {title} {Optical spectrum of mos2: Many-body effects and diversity of exciton
  states},}\ }\href@noop {} {\bibfield  {journal} {\bibinfo  {journal} {Phys.
  Rev. Lett.}\ }\textbf {\bibinfo {volume} {111}},\ \bibinfo {pages} {216805}
  (\bibinfo {year} {2013})}\BibitemShut {NoStop}%
\bibitem [{\citenamefont {Ramasubramaniam}(2012)}]{Ramasubramaniam:2012a}%
  \BibitemOpen
  \bibfield  {author} {\bibinfo {author} {\bibfnamefont {Ashwin}\ \bibnamefont
  {Ramasubramaniam}},\ }\bibfield  {title} {\enquote {\bibinfo {title} {Large
  excitonic effects in monolayers of molybdenum and tungsten
  dichalcogenides},}\ }\href@noop {} {\bibfield  {journal} {\bibinfo  {journal}
  {Phys. Rev. B}\ }\textbf {\bibinfo {volume} {86}},\ \bibinfo {pages} {115409}
  (\bibinfo {year} {2012})}\BibitemShut {NoStop}%
\bibitem [{\citenamefont {Wang}\ \emph {et~al.}(2015)\citenamefont {Wang},
  \citenamefont {Marie}, \citenamefont {Gerber}, \citenamefont {Amand},
  \citenamefont {Lagarde}, \citenamefont {Bouet}, \citenamefont {Vidal},
  \citenamefont {Balocchi},\ and\ \citenamefont {Urbaszek}}]{Wang:2015b}%
  \BibitemOpen
  \bibfield  {author} {\bibinfo {author} {\bibfnamefont {G.}~\bibnamefont
  {Wang}}, \bibinfo {author} {\bibfnamefont {X.}~\bibnamefont {Marie}},
  \bibinfo {author} {\bibfnamefont {I.}~\bibnamefont {Gerber}}, \bibinfo
  {author} {\bibfnamefont {T.}~\bibnamefont {Amand}}, \bibinfo {author}
  {\bibfnamefont {D.}~\bibnamefont {Lagarde}}, \bibinfo {author} {\bibfnamefont
  {L.}~\bibnamefont {Bouet}}, \bibinfo {author} {\bibfnamefont
  {M.}~\bibnamefont {Vidal}}, \bibinfo {author} {\bibfnamefont
  {A.}~\bibnamefont {Balocchi}}, \ and\ \bibinfo {author} {\bibfnamefont
  {B.}~\bibnamefont {Urbaszek}},\ }\bibfield  {title} {\enquote {\bibinfo
  {title} {Giant enhancement of the optical second-harmonic emission of
  ${\mathrm{wse}}_{2}$ monolayers by laser excitation at exciton resonances},}\
  }\href@noop {} {\bibfield  {journal} {\bibinfo  {journal} {Phys. Rev. Lett.}\
  }\textbf {\bibinfo {volume} {114}},\ \bibinfo {pages} {097403} (\bibinfo
  {year} {2015})}\BibitemShut {NoStop}%
\bibitem [{\citenamefont {Xiao}\ \emph {et~al.}(2012)\citenamefont {Xiao},
  \citenamefont {Liu}, \citenamefont {Feng}, \citenamefont {Xu},\ and\
  \citenamefont {Yao}}]{Xiao:2012a}%
  \BibitemOpen
  \bibfield  {author} {\bibinfo {author} {\bibfnamefont {Di}~\bibnamefont
  {Xiao}}, \bibinfo {author} {\bibfnamefont {Gui-Bin}\ \bibnamefont {Liu}},
  \bibinfo {author} {\bibfnamefont {Wanxiang}\ \bibnamefont {Feng}}, \bibinfo
  {author} {\bibfnamefont {Xiaodong}\ \bibnamefont {Xu}}, \ and\ \bibinfo
  {author} {\bibfnamefont {Wang}\ \bibnamefont {Yao}},\ }\bibfield  {title}
  {\enquote {\bibinfo {title} {Coupled spin and valley physics in monolayers of
  ${\mathrm{mos}}_{2}$ and other group-vi dichalcogenides},}\ }\href@noop {}
  {\bibfield  {journal} {\bibinfo  {journal} {Phys. Rev. Lett.}\ }\textbf
  {\bibinfo {volume} {108}},\ \bibinfo {pages} {196802} (\bibinfo {year}
  {2012})}\BibitemShut {NoStop}%
\bibitem [{\citenamefont {Sallen}\ \emph {et~al.}(2012)\citenamefont {Sallen},
  \citenamefont {Bouet}, \citenamefont {Marie}, \citenamefont {Wang},
  \citenamefont {Zhu}, \citenamefont {Han}, \citenamefont {Lu}, \citenamefont
  {Tan}, \citenamefont {Amand}, \citenamefont {Liu},\ and\ \citenamefont
  {Urbaszek}}]{Sallen:2012a}%
  \BibitemOpen
  \bibfield  {author} {\bibinfo {author} {\bibfnamefont {G.}~\bibnamefont
  {Sallen}}, \bibinfo {author} {\bibfnamefont {L.}~\bibnamefont {Bouet}},
  \bibinfo {author} {\bibfnamefont {X.}~\bibnamefont {Marie}}, \bibinfo
  {author} {\bibfnamefont {G.}~\bibnamefont {Wang}}, \bibinfo {author}
  {\bibfnamefont {C.~R.}\ \bibnamefont {Zhu}}, \bibinfo {author} {\bibfnamefont
  {W.~P.}\ \bibnamefont {Han}}, \bibinfo {author} {\bibfnamefont
  {Y.}~\bibnamefont {Lu}}, \bibinfo {author} {\bibfnamefont {P.~H.}\
  \bibnamefont {Tan}}, \bibinfo {author} {\bibfnamefont {T.}~\bibnamefont
  {Amand}}, \bibinfo {author} {\bibfnamefont {B.~L.}\ \bibnamefont {Liu}}, \
  and\ \bibinfo {author} {\bibfnamefont {B.}~\bibnamefont {Urbaszek}},\
  }\bibfield  {title} {\enquote {\bibinfo {title} {Robust optical emission
  polarization in mos${}_{2}$ monolayers through selective valley
  excitation},}\ }\href@noop {} {\bibfield  {journal} {\bibinfo  {journal}
  {Phys. Rev. B}\ }\textbf {\bibinfo {volume} {86}},\ \bibinfo {pages} {081301}
  (\bibinfo {year} {2012})}\BibitemShut {NoStop}%
\bibitem [{\citenamefont {Mak}\ \emph {et~al.}(2012)\citenamefont {Mak},
  \citenamefont {He}, \citenamefont {Shan},\ and\ \citenamefont
  {Heinz}}]{Mak:2012a}%
  \BibitemOpen
  \bibfield  {author} {\bibinfo {author} {\bibfnamefont {Kin~Fai}\ \bibnamefont
  {Mak}}, \bibinfo {author} {\bibfnamefont {Keliang}\ \bibnamefont {He}},
  \bibinfo {author} {\bibfnamefont {Jie}\ \bibnamefont {Shan}}, \ and\ \bibinfo
  {author} {\bibfnamefont {Tony~F.}\ \bibnamefont {Heinz}},\ }\bibfield
  {title} {\enquote {\bibinfo {title} {Control of valley polarization in
  monolayer ${\mathrm{mos}}_{2}$ by optical helicity},}\ }\href@noop {}
  {\bibfield  {journal} {\bibinfo  {journal} {Nat. Nanotechnol.}\ }\textbf
  {\bibinfo {volume} {7}},\ \bibinfo {pages} {494} (\bibinfo {year}
  {2012})}\BibitemShut {NoStop}%
\bibitem [{\citenamefont {Kioseoglou}\ \emph {et~al.}(2012)\citenamefont
  {Kioseoglou}, \citenamefont {Hanbicki}, \citenamefont {Currie}, \citenamefont
  {Friedman}, \citenamefont {Gunlycke},\ and\ \citenamefont
  {Jonker}}]{Kioseoglou:2012a}%
  \BibitemOpen
  \bibfield  {author} {\bibinfo {author} {\bibfnamefont {G.}~\bibnamefont
  {Kioseoglou}}, \bibinfo {author} {\bibfnamefont {A.~T.}\ \bibnamefont
  {Hanbicki}}, \bibinfo {author} {\bibfnamefont {M.}~\bibnamefont {Currie}},
  \bibinfo {author} {\bibfnamefont {A.~L.}\ \bibnamefont {Friedman}}, \bibinfo
  {author} {\bibfnamefont {D.}~\bibnamefont {Gunlycke}}, \ and\ \bibinfo
  {author} {\bibfnamefont {B.~T.}\ \bibnamefont {Jonker}},\ }\bibfield  {title}
  {\enquote {\bibinfo {title} {Valley polarization and intervalley scattering
  in monolayer mos[sub 2]},}\ }\href@noop {} {\bibfield  {journal} {\bibinfo
  {journal} {Applied Physics Letters}\ }\textbf {\bibinfo {volume} {101}},\
  \bibinfo {eid} {221907} (\bibinfo {year} {2012})}\BibitemShut {NoStop}%
\bibitem [{\citenamefont {Cao}\ \emph {et~al.}(2012)\citenamefont {Cao},
  \citenamefont {Wang}, \citenamefont {Han}, \citenamefont {Ye}, \citenamefont
  {Zhu}, \citenamefont {Shi}, \citenamefont {Niu}, \citenamefont {Tan},
  \citenamefont {Wang}, \citenamefont {Liu},\ and\ \citenamefont
  {Feng}}]{Cao:2012a}%
  \BibitemOpen
  \bibfield  {author} {\bibinfo {author} {\bibfnamefont {Ting}\ \bibnamefont
  {Cao}}, \bibinfo {author} {\bibfnamefont {Gang}\ \bibnamefont {Wang}},
  \bibinfo {author} {\bibfnamefont {Wenpeng}\ \bibnamefont {Han}}, \bibinfo
  {author} {\bibfnamefont {Huiqui}\ \bibnamefont {Ye}}, \bibinfo {author}
  {\bibfnamefont {Chuanrui}\ \bibnamefont {Zhu}}, \bibinfo {author}
  {\bibfnamefont {Junren}\ \bibnamefont {Shi}}, \bibinfo {author}
  {\bibfnamefont {Qian}\ \bibnamefont {Niu}}, \bibinfo {author} {\bibfnamefont
  {Pingheng}\ \bibnamefont {Tan}}, \bibinfo {author} {\bibfnamefont {Enge}\
  \bibnamefont {Wang}}, \bibinfo {author} {\bibfnamefont {Baoli}\ \bibnamefont
  {Liu}}, \ and\ \bibinfo {author} {\bibfnamefont {Ji}~\bibnamefont {Feng}},\
  }\bibfield  {title} {\enquote {\bibinfo {title} {Valley-selective circular
  dichroism in ${\mathrm{mos}}_{2}$},}\ }\href@noop {} {\bibfield  {journal}
  {\bibinfo  {journal} {Nature Communications}\ }\textbf {\bibinfo {volume}
  {3}},\ \bibinfo {pages} {887} (\bibinfo {year} {2012})}\BibitemShut {NoStop}%
\bibitem [{\citenamefont {Jones}\ \emph {et~al.}(2013)\citenamefont {Jones},
  \citenamefont {Yu}, \citenamefont {Ghimire}, \citenamefont {Wu},
  \citenamefont {Aivazian}, \citenamefont {Ross}, \citenamefont {Zhao},
  \citenamefont {Yan}, \citenamefont {Mandrus}, \citenamefont {Xiao},
  \citenamefont {Yao},\ and\ \citenamefont {Xu}}]{Jones:2013a}%
  \BibitemOpen
  \bibfield  {author} {\bibinfo {author} {\bibfnamefont {Aaron~M.}\
  \bibnamefont {Jones}}, \bibinfo {author} {\bibfnamefont {Hongyi}\
  \bibnamefont {Yu}}, \bibinfo {author} {\bibfnamefont {Nirmal~J.}\
  \bibnamefont {Ghimire}}, \bibinfo {author} {\bibfnamefont {Sanfeng}\
  \bibnamefont {Wu}}, \bibinfo {author} {\bibfnamefont {Grant}\ \bibnamefont
  {Aivazian}}, \bibinfo {author} {\bibfnamefont {Jason~S.}\ \bibnamefont
  {Ross}}, \bibinfo {author} {\bibfnamefont {Bo}~\bibnamefont {Zhao}}, \bibinfo
  {author} {\bibfnamefont {Jiaqiang}\ \bibnamefont {Yan}}, \bibinfo {author}
  {\bibfnamefont {David~G.}\ \bibnamefont {Mandrus}}, \bibinfo {author}
  {\bibfnamefont {Di}~\bibnamefont {Xiao}}, \bibinfo {author} {\bibfnamefont
  {Wang}\ \bibnamefont {Yao}}, \ and\ \bibinfo {author} {\bibfnamefont
  {Xiaodong}\ \bibnamefont {Xu}},\ }\bibfield  {title} {\enquote {\bibinfo
  {title} {Optical generation of excitonic valley coherence in monolayer
  wse2},}\ }\href@noop {} {\bibfield  {journal} {\bibinfo  {journal} {Nat.
  Nanotechnol.}\ }\textbf {\bibinfo {volume} {8}},\ \bibinfo {pages} {634--638}
  (\bibinfo {year} {2013})}\BibitemShut {NoStop}%
\bibitem [{\citenamefont {Yang}\ \emph {et~al.}(2015)\citenamefont {Yang},
  \citenamefont {Sinitsyn}, \citenamefont {Chen}, \citenamefont {Yuan},
  \citenamefont {Zhang}, \citenamefont {Lou},\ and\ \citenamefont
  {Crooker}}]{Yang:2015a}%
  \BibitemOpen
  \bibfield  {author} {\bibinfo {author} {\bibfnamefont {Luyi}\ \bibnamefont
  {Yang}}, \bibinfo {author} {\bibfnamefont {Nikolai~A}\ \bibnamefont
  {Sinitsyn}}, \bibinfo {author} {\bibfnamefont {Weibing}\ \bibnamefont
  {Chen}}, \bibinfo {author} {\bibfnamefont {Jiangtan}\ \bibnamefont {Yuan}},
  \bibinfo {author} {\bibfnamefont {Jing}\ \bibnamefont {Zhang}}, \bibinfo
  {author} {\bibfnamefont {Jun}\ \bibnamefont {Lou}}, \ and\ \bibinfo {author}
  {\bibfnamefont {Scott~A}\ \bibnamefont {Crooker}},\ }\bibfield  {title}
  {\enquote {\bibinfo {title} {Long-lived nanosecond spin relaxation and spin
  coherence of electrons in monolayer mos2 and ws2},}\ }\href@noop {}
  {\bibfield  {journal} {\bibinfo  {journal} {Nature Physics}\ }\textbf
  {\bibinfo {volume} {11}},\ \bibinfo {pages} {830--834} (\bibinfo {year}
  {2015})}\BibitemShut {NoStop}%
\bibitem [{\citenamefont {Weiner}\ \emph {et~al.}(1985)\citenamefont {Weiner},
  \citenamefont {Chemla}, \citenamefont {Miller}, \citenamefont {Haus},
  \citenamefont {Gossard}, \citenamefont {Wiegmann},\ and\ \citenamefont
  {Burrus}}]{Miller:1985}%
  \BibitemOpen
  \bibfield  {author} {\bibinfo {author} {\bibfnamefont {J.~S.}\ \bibnamefont
  {Weiner}}, \bibinfo {author} {\bibfnamefont {D.~S.}\ \bibnamefont {Chemla}},
  \bibinfo {author} {\bibfnamefont {D.~A.~B.}\ \bibnamefont {Miller}}, \bibinfo
  {author} {\bibfnamefont {H.~A.}\ \bibnamefont {Haus}}, \bibinfo {author}
  {\bibfnamefont {A.~C.}\ \bibnamefont {Gossard}}, \bibinfo {author}
  {\bibfnamefont {W.}~\bibnamefont {Wiegmann}}, \ and\ \bibinfo {author}
  {\bibfnamefont {C.~A.}\ \bibnamefont {Burrus}},\ }\bibfield  {title}
  {\enquote {\bibinfo {title} {Highly anisotropic optical properties of single
  quantum well waveguides},}\ }\href {\doibase 10.1063/1.96051} {\bibfield
  {journal} {\bibinfo  {journal} {Applied Physics Letters}\ }\textbf {\bibinfo
  {volume} {47}},\ \bibinfo {pages} {664--667} (\bibinfo {year}
  {1985})}\BibitemShut {NoStop}%
\bibitem [{\citenamefont {Marzin}\ \emph {et~al.}(1985)\citenamefont {Marzin},
  \citenamefont {Charasse},\ and\ \citenamefont {Sermage}}]{Marzin:1985}%
  \BibitemOpen
  \bibfield  {author} {\bibinfo {author} {\bibfnamefont {J.-Y.}\ \bibnamefont
  {Marzin}}, \bibinfo {author} {\bibfnamefont {M.~N.}\ \bibnamefont
  {Charasse}}, \ and\ \bibinfo {author} {\bibfnamefont {B.}~\bibnamefont
  {Sermage}},\ }\bibfield  {title} {\enquote {\bibinfo {title} {Optical
  investigation of a new type of valence-band configuration in
  ${\mathrm{in}}_{\mathrm{x}}$ ${\mathrm{ga}}_{1\mathrm{\ensuremath{-}}\mathrm{x}}$as-gaas
  strained superlattices},}\ }\href {\doibase 10.1103/PhysRevB.31.8298}
  {\bibfield  {journal} {\bibinfo  {journal} {Phys. Rev. B}\ }\textbf {\bibinfo
  {volume} {31}},\ \bibinfo {pages} {8298--8301} (\bibinfo {year}
  {1985})}\BibitemShut {NoStop}%
\bibitem [{\citenamefont {Glazov}\ \emph {et~al.}(2014)\citenamefont {Glazov},
  \citenamefont {Amand}, \citenamefont {Marie}, \citenamefont {Lagarde},
  \citenamefont {Bouet},\ and\ \citenamefont {Urbaszek}}]{Glazov:2014a}%
  \BibitemOpen
  \bibfield  {author} {\bibinfo {author} {\bibfnamefont {M.~M.}\ \bibnamefont
  {Glazov}}, \bibinfo {author} {\bibfnamefont {T.}~\bibnamefont {Amand}},
  \bibinfo {author} {\bibfnamefont {X.}~\bibnamefont {Marie}}, \bibinfo
  {author} {\bibfnamefont {D.}~\bibnamefont {Lagarde}}, \bibinfo {author}
  {\bibfnamefont {L.}~\bibnamefont {Bouet}}, \ and\ \bibinfo {author}
  {\bibfnamefont {B.}~\bibnamefont {Urbaszek}},\ }\bibfield  {title} {\enquote
  {\bibinfo {title} {Exciton fine structure and spin decoherence in monolayers
  of transition metal dichalcogenides},}\ }\href@noop {} {\bibfield  {journal}
  {\bibinfo  {journal} {Phys. Rev. B}\ }\textbf {\bibinfo {volume} {89}},\
  \bibinfo {pages} {201302} (\bibinfo {year} {2014})}\BibitemShut {NoStop}%
\bibitem [{\citenamefont {Echeverry}\ \emph {et~al.}(2016)\citenamefont
  {Echeverry}, \citenamefont {Urbaszek}, \citenamefont {Amand}, \citenamefont
  {Marie},\ and\ \citenamefont {Gerber}}]{Echeverry:2016}%
  \BibitemOpen
  \bibfield  {author} {\bibinfo {author} {\bibfnamefont {J.~P.}\ \bibnamefont
  {Echeverry}}, \bibinfo {author} {\bibfnamefont {B.}~\bibnamefont {Urbaszek}},
  \bibinfo {author} {\bibfnamefont {T.}~\bibnamefont {Amand}}, \bibinfo
  {author} {\bibfnamefont {X.}~\bibnamefont {Marie}}, \ and\ \bibinfo {author}
  {\bibfnamefont {I.~C.}\ \bibnamefont {Gerber}},\ }\bibfield  {title}
  {\enquote {\bibinfo {title} {Splitting between bright and dark excitons in
  transition metal dichalcogenide monolayers},}\ }\href {\doibase
  10.1103/PhysRevB.93.121107} {\bibfield  {journal} {\bibinfo  {journal} {Phys.
  Rev. B}\ }\textbf {\bibinfo {volume} {93}},\ \bibinfo {pages} {121107}
  (\bibinfo {year} {2016})}\BibitemShut {NoStop}%
\bibitem [{\citenamefont {Zhang}\ \emph {et~al.}(2015)\citenamefont {Zhang},
  \citenamefont {You}, \citenamefont {Zhao},\ and\ \citenamefont
  {Heinz}}]{Zhang:2015d}%
  \BibitemOpen
  \bibfield  {author} {\bibinfo {author} {\bibfnamefont {Xiao-Xiao}\
  \bibnamefont {Zhang}}, \bibinfo {author} {\bibfnamefont {Yumeng}\
  \bibnamefont {You}}, \bibinfo {author} {\bibfnamefont {Shu Yang~Frank}\
  \bibnamefont {Zhao}}, \ and\ \bibinfo {author} {\bibfnamefont {Tony~F.}\
  \bibnamefont {Heinz}},\ }\bibfield  {title} {\enquote {\bibinfo {title}
  {Experimental evidence for dark excitons in monolayer
  ${\mathrm{wse}}_{2}$},}\ }\href {\doibase 10.1103/PhysRevLett.115.257403}
  {\bibfield  {journal} {\bibinfo  {journal} {Phys. Rev. Lett.}\ }\textbf
  {\bibinfo {volume} {115}},\ \bibinfo {pages} {257403} (\bibinfo {year}
  {2015})}\BibitemShut {NoStop}%
\bibitem [{\citenamefont {{Wang}}\ \emph {et~al.}(2015)\citenamefont {{Wang}},
  \citenamefont {{Robert}}, \citenamefont {{Suslu}}, \citenamefont {{Chen}},
  \citenamefont {{Yang}}, \citenamefont {{Alamdari}}, \citenamefont {{Gerber}},
  \citenamefont {{Amand}}, \citenamefont {{Marie}}, \citenamefont {{Tongay}},\
  and\ \citenamefont {{Urbaszek}}}]{Wang:2015e}%
  \BibitemOpen
  \bibfield  {author} {\bibinfo {author} {\bibfnamefont {G.}~\bibnamefont
  {{Wang}}}, \bibinfo {author} {\bibfnamefont {C.}~\bibnamefont {{Robert}}},
  \bibinfo {author} {\bibfnamefont {A.}~\bibnamefont {{Suslu}}}, \bibinfo
  {author} {\bibfnamefont {B.}~\bibnamefont {{Chen}}}, \bibinfo {author}
  {\bibfnamefont {S.}~\bibnamefont {{Yang}}}, \bibinfo {author} {\bibfnamefont
  {S.}~\bibnamefont {{Alamdari}}}, \bibinfo {author} {\bibfnamefont {I.~C.}\
  \bibnamefont {{Gerber}}}, \bibinfo {author} {\bibfnamefont {T.}~\bibnamefont
  {{Amand}}}, \bibinfo {author} {\bibfnamefont {X.}~\bibnamefont {{Marie}}},
  \bibinfo {author} {\bibfnamefont {S.}~\bibnamefont {{Tongay}}}, \ and\
  \bibinfo {author} {\bibfnamefont {B.}~\bibnamefont {{Urbaszek}}},\ }\bibfield
   {title} {\enquote {\bibinfo {title} {{Spin-orbit engineering in transition
  metal dichalcogenide alloy monolayers}},}\ }\href@noop {} {\bibfield
  {journal} {\bibinfo  {journal} {Nature Comms.}\ ,\ \bibinfo {pages} {10110}}
  (\bibinfo {year} {2015})}\BibitemShut {NoStop}%
\bibitem [{\citenamefont {Arora}\ \emph {et~al.}(2015)\citenamefont {Arora},
  \citenamefont {Koperski}, \citenamefont {Nogajewski}, \citenamefont {Marcus},
  \citenamefont {Faugeras},\ and\ \citenamefont {Potemski}}]{Arora:2015a}%
  \BibitemOpen
  \bibfield  {author} {\bibinfo {author} {\bibfnamefont {Ashish}\ \bibnamefont
  {Arora}}, \bibinfo {author} {\bibfnamefont {Maciej}\ \bibnamefont
  {Koperski}}, \bibinfo {author} {\bibfnamefont {Karol}\ \bibnamefont
  {Nogajewski}}, \bibinfo {author} {\bibfnamefont {Jacques}\ \bibnamefont
  {Marcus}}, \bibinfo {author} {\bibfnamefont {Clement}\ \bibnamefont
  {Faugeras}}, \ and\ \bibinfo {author} {\bibfnamefont {Marek}\ \bibnamefont
  {Potemski}},\ }\bibfield  {title} {\enquote {\bibinfo {title} {Excitonic
  resonances in thin films of wse2: from monolayer to bulk material},}\
  }\href@noop {} {\bibfield  {journal} {\bibinfo  {journal} {Nanoscale}\
  }\textbf {\bibinfo {volume} {7}},\ \bibinfo {pages} {10421--10429} (\bibinfo
  {year} {2015})}\BibitemShut {NoStop}%
\bibitem [{\citenamefont {Withers}\ \emph {et~al.}(2015)\citenamefont
  {Withers}, \citenamefont {Del Pozo-Zamudio}, \citenamefont {Schwarz},
  \citenamefont {Dufferwiel}, \citenamefont {Walker}, \citenamefont {Godde},
  \citenamefont {Rooney}, \citenamefont {Gholinia}, \citenamefont {Woods},
  \citenamefont {Blake}, \citenamefont {Haigh}, \citenamefont {Watanabe},
  \citenamefont {Taniguchi}, \citenamefont {Aleiner}, \citenamefont {Geim},
  \citenamefont {Fal???ko}, \citenamefont {Tartakovskii},\ and\
  \citenamefont {Novoselov}}]{Withers:2015}%
  \BibitemOpen
  \bibfield  {author} {\bibinfo {author} {\bibfnamefont {F.}~\bibnamefont
  {Withers}}, \bibinfo {author} {\bibfnamefont {O.}~\bibnamefont {Del
  Pozo-Zamudio}}, \bibinfo {author} {\bibfnamefont {S.}~\bibnamefont
  {Schwarz}}, \bibinfo {author} {\bibfnamefont {S.}~\bibnamefont {Dufferwiel}},
  \bibinfo {author} {\bibfnamefont {P.~M.}\ \bibnamefont {Walker}}, \bibinfo
  {author} {\bibfnamefont {T.}~\bibnamefont {Godde}}, \bibinfo {author}
  {\bibfnamefont {A.~P.}\ \bibnamefont {Rooney}}, \bibinfo {author}
  {\bibfnamefont {A.}~\bibnamefont {Gholinia}}, \bibinfo {author}
  {\bibfnamefont {C.~R.}\ \bibnamefont {Woods}}, \bibinfo {author}
  {\bibfnamefont {P.}~\bibnamefont {Blake}}, \bibinfo {author} {\bibfnamefont
  {S.~J.}\ \bibnamefont {Haigh}}, \bibinfo {author} {\bibfnamefont
  {K.}~\bibnamefont {Watanabe}}, \bibinfo {author} {\bibfnamefont
  {T.}~\bibnamefont {Taniguchi}}, \bibinfo {author} {\bibfnamefont {I.~L.}\
  \bibnamefont {Aleiner}}, \bibinfo {author} {\bibfnamefont {A.~K.}\
  \bibnamefont {Geim}}, \bibinfo {author} {\bibfnamefont {V.~I.}\ \bibnamefont
  {Fal???ko}}, \bibinfo {author} {\bibfnamefont {A.~I.}\ \bibnamefont
  {Tartakovskii}}, \ and\ \bibinfo {author} {\bibfnamefont {K.~S.}\
  \bibnamefont {Novoselov}},\ }\bibfield  {title} {\enquote {\bibinfo {title}
  {Wse2 light-emitting tunneling transistors with enhanced brightness at room
  temperature},}\ }\href {\doibase 10.1021/acs.nanolett.5b03740} {\bibfield
  {journal} {\bibinfo  {journal} {Nano Letters}\ }\textbf {\bibinfo {volume}
  {15}},\ \bibinfo {pages} {8223--8228} (\bibinfo {year} {2015})},\ \bibinfo
  {note} {pMID: 26555037},\ \Eprint
  {http://arxiv.org/abs/http://dx.doi.org/10.1021/acs.nanolett.5b03740}
  {http://dx.doi.org/10.1021/acs.nanolett.5b03740} \BibitemShut {NoStop}%
\bibitem [{\citenamefont {Ko\ifmmode~\acute{s}\else \'{s}\fi{}mider}\ \emph
  {et~al.}(2013)\citenamefont {Ko\ifmmode~\acute{s}\else \'{s}\fi{}mider},
  \citenamefont {Gonz\'alez},\ and\ \citenamefont
  {Fern\'andez-Rossier}}]{Kosmider:2013b}%
  \BibitemOpen
  \bibfield  {author} {\bibinfo {author} {\bibfnamefont {K.}~\bibnamefont
  {Ko\ifmmode~\acute{s}\else \'{s}\fi{}mider}}, \bibinfo {author}
  {\bibfnamefont {J.~W.}\ \bibnamefont {Gonz\'alez}}, \ and\ \bibinfo {author}
  {\bibfnamefont {J.}~\bibnamefont {Fern\'andez-Rossier}},\ }\bibfield  {title}
  {\enquote {\bibinfo {title} {Large spin splitting in the conduction band of
  transition metal dichalcogenide monolayers},}\ }\href {\doibase
  10.1103/PhysRevB.88.245436} {\bibfield  {journal} {\bibinfo  {journal} {Phys.
  Rev. B}\ }\textbf {\bibinfo {volume} {88}},\ \bibinfo {pages} {245436}
  (\bibinfo {year} {2013})}\BibitemShut {NoStop}%
\bibitem [{\citenamefont {{Kormanyos}}\ \emph {et~al.}(2015)\citenamefont
  {{Kormanyos}}, \citenamefont {{Burkard}}, \citenamefont {{Gmitra}},
  \citenamefont {{Fabian}}, \citenamefont {{Zolyomi}}, \citenamefont
  {{Drummond}},\ and\ \citenamefont {{Fal'ko}}}]{Kormanyos:2015a}%
  \BibitemOpen
  \bibfield  {author} {\bibinfo {author} {\bibfnamefont {A.}~\bibnamefont
  {{Kormanyos}}}, \bibinfo {author} {\bibfnamefont {G.}~\bibnamefont
  {{Burkard}}}, \bibinfo {author} {\bibfnamefont {M.}~\bibnamefont {{Gmitra}}},
  \bibinfo {author} {\bibfnamefont {J.}~\bibnamefont {{Fabian}}}, \bibinfo
  {author} {\bibfnamefont {V.}~\bibnamefont {{Zolyomi}}}, \bibinfo {author}
  {\bibfnamefont {N.~D.}\ \bibnamefont {{Drummond}}}, \ and\ \bibinfo {author}
  {\bibfnamefont {V.}~\bibnamefont {{Fal'ko}}},\ }\bibfield  {title} {\enquote
  {\bibinfo {title} {{k.p theory for two-dimensional transition metal
  dichalcogenide semiconductors}},}\ }\href@noop {} {\bibfield  {journal}
  {\bibinfo  {journal} {2D Materials}\ }\textbf {\bibinfo {volume} {2}},\
  \bibinfo {pages} {022001} (\bibinfo {year} {2015})}\BibitemShut {NoStop}%
\bibitem [{\citenamefont {Dery}\ and\ \citenamefont {Song}(2015)}]{Dery:2015a}%
  \BibitemOpen
  \bibfield  {author} {\bibinfo {author} {\bibfnamefont {Hanan}\ \bibnamefont
  {Dery}}\ and\ \bibinfo {author} {\bibfnamefont {Yang}\ \bibnamefont {Song}},\
  }\bibfield  {title} {\enquote {\bibinfo {title} {Polarization analysis of
  excitons in monolayer and bilayer transition-metal dichalcogenides},}\
  }\href@noop {} {\bibfield  {journal} {\bibinfo  {journal} {Phys. Rev. B}\
  }\textbf {\bibinfo {volume} {92}},\ \bibinfo {pages} {125431} (\bibinfo
  {year} {2015})}\BibitemShut {NoStop}%
\bibitem [{\citenamefont {Wang}\ \emph {et~al.}(2017)\citenamefont {Wang},
  \citenamefont {Zhao}, \citenamefont {Mak},\ and\ \citenamefont
  {Shan}}]{Wang:2017b}%
  \BibitemOpen
  \bibfield  {author} {\bibinfo {author} {\bibfnamefont {Zefang}\ \bibnamefont
  {Wang}}, \bibinfo {author} {\bibfnamefont {Liang}\ \bibnamefont {Zhao}},
  \bibinfo {author} {\bibfnamefont {Kin~Fai}\ \bibnamefont {Mak}}, \ and\
  \bibinfo {author} {\bibfnamefont {Jie}\ \bibnamefont {Shan}},\ }\bibfield
  {title} {\enquote {\bibinfo {title} {Probing the spin-polarized electronic
  band structure in monolayer transition metal dichalcogenides by optical
  spectroscopy},}\ }\href {\doibase 10.1021/acs.nanolett.6b03855} {\bibfield
  {journal} {\bibinfo  {journal} {Nano Letters}\ }\textbf {\bibinfo {volume}
  {17}},\ \bibinfo {pages} {740--746} (\bibinfo {year} {2017})},\ \bibinfo
  {note} {pMID: 28103668},\ \Eprint
  {http://arxiv.org/abs/http://dx.doi.org/10.1021/acs.nanolett.6b03855}
  {http://dx.doi.org/10.1021/acs.nanolett.6b03855} \BibitemShut {NoStop}%
\bibitem [{\citenamefont {{Ajayi}}\ \emph {et~al.}(2017)\citenamefont
  {{Ajayi}}, \citenamefont {{Ardelean}}, \citenamefont {{Shepard}},
  \citenamefont {{Wang}}, \citenamefont {{Antony}}, \citenamefont
  {{Taniguchi}}, \citenamefont {{Watanabe}}, \citenamefont {{Heinz}},
  \citenamefont {{Strauf}}, \citenamefont {{Zhu}},\ and\ \citenamefont
  {{Hone}}}]{Ajayi:2017}%
  \BibitemOpen
  \bibfield  {author} {\bibinfo {author} {\bibfnamefont {O.~A.}\ \bibnamefont
  {{Ajayi}}}, \bibinfo {author} {\bibfnamefont {J.~V.}\ \bibnamefont
  {{Ardelean}}}, \bibinfo {author} {\bibfnamefont {G.~D.}\ \bibnamefont
  {{Shepard}}}, \bibinfo {author} {\bibfnamefont {J.}~\bibnamefont {{Wang}}},
  \bibinfo {author} {\bibfnamefont {A.}~\bibnamefont {{Antony}}}, \bibinfo
  {author} {\bibfnamefont {T.}~\bibnamefont {{Taniguchi}}}, \bibinfo {author}
  {\bibfnamefont {K.}~\bibnamefont {{Watanabe}}}, \bibinfo {author}
  {\bibfnamefont {T.~F.}\ \bibnamefont {{Heinz}}}, \bibinfo {author}
  {\bibfnamefont {S.}~\bibnamefont {{Strauf}}}, \bibinfo {author}
  {\bibfnamefont {X.-Y.}\ \bibnamefont {{Zhu}}}, \ and\ \bibinfo {author}
  {\bibfnamefont {J.~C.}\ \bibnamefont {{Hone}}},\ }\bibfield  {title}
  {\enquote {\bibinfo {title} {{Approaching the Intrinsic Photoluminescence
  Linewidth in Transition Metal Dichalcogenide Monolayers}},}\ }\href@noop {}
  {\bibfield  {journal} {\bibinfo  {journal} {ArXiv e-prints}\ } (\bibinfo
  {year} {2017})},\ \Eprint {http://arxiv.org/abs/1702.05857} {arXiv:1702.05857
  [cond-mat.mtrl-sci]} \BibitemShut {NoStop}%
\bibitem [{\citenamefont {{Cadiz}}\ \emph {et~al.}(2017)\citenamefont
  {{Cadiz}}, \citenamefont {{Courtade}}, \citenamefont {{Robert}},
  \citenamefont {{Wang}}, \citenamefont {{Shen}}, \citenamefont {{Cai}},
  \citenamefont {{Taniguchi}}, \citenamefont {{Watanabe}}, \citenamefont
  {{Carrere}}, \citenamefont {{Lagarde}}, \citenamefont {{Manca}},
  \citenamefont {{Amand}}, \citenamefont {{Renucci}}, \citenamefont {{Tongay}},
  \citenamefont {{Marie}},\ and\ \citenamefont {{Urbaszek}}}]{Cadiz:2017a}%
  \BibitemOpen
  \bibfield  {author} {\bibinfo {author} {\bibfnamefont {F.}~\bibnamefont
  {{Cadiz}}}, \bibinfo {author} {\bibfnamefont {E.}~\bibnamefont {{Courtade}}},
  \bibinfo {author} {\bibfnamefont {C.}~\bibnamefont {{Robert}}}, \bibinfo
  {author} {\bibfnamefont {G.}~\bibnamefont {{Wang}}}, \bibinfo {author}
  {\bibfnamefont {Y.}~\bibnamefont {{Shen}}}, \bibinfo {author} {\bibfnamefont
  {H.}~\bibnamefont {{Cai}}}, \bibinfo {author} {\bibfnamefont
  {T.}~\bibnamefont {{Taniguchi}}}, \bibinfo {author} {\bibfnamefont
  {K.}~\bibnamefont {{Watanabe}}}, \bibinfo {author} {\bibfnamefont
  {H.}~\bibnamefont {{Carrere}}}, \bibinfo {author} {\bibfnamefont
  {D.}~\bibnamefont {{Lagarde}}}, \bibinfo {author} {\bibfnamefont
  {M.}~\bibnamefont {{Manca}}}, \bibinfo {author} {\bibfnamefont
  {T.}~\bibnamefont {{Amand}}}, \bibinfo {author} {\bibfnamefont
  {P.}~\bibnamefont {{Renucci}}}, \bibinfo {author} {\bibfnamefont
  {S.}~\bibnamefont {{Tongay}}}, \bibinfo {author} {\bibfnamefont
  {X.}~\bibnamefont {{Marie}}}, \ and\ \bibinfo {author} {\bibfnamefont
  {B.}~\bibnamefont {{Urbaszek}}},\ }\bibfield  {title} {\enquote {\bibinfo
  {title} {{Excitonic linewidth approaching the homogeneous limit in MoS2 based
  van der Waals heterostructures : accessing spin-valley dynamics}},}\
  }\href@noop {} {\bibfield  {journal} {\bibinfo  {journal} {ArXiv e-prints}\ }
  (\bibinfo {year} {2017})},\ \Eprint {http://arxiv.org/abs/1702.00323}
  {arXiv:1702.00323 [cond-mat.mtrl-sci]} \BibitemShut {NoStop}%
\bibitem [{\citenamefont {Manca}\ \emph {et~al.}(2016)\citenamefont {Manca},
  \citenamefont {Glazov}, \citenamefont {Robert}, \citenamefont {Cadiz},
  \citenamefont {Taniguchi}, \citenamefont {Watanabe}, \citenamefont
  {Courtade}, \citenamefont {Amand}, \citenamefont {Renucci}, \citenamefont
  {Marie}, \citenamefont {Wang},\ and\ \citenamefont {Urbaszek}}]{Manca:2017a}%
  \BibitemOpen
  \bibfield  {author} {\bibinfo {author} {\bibfnamefont {M}~\bibnamefont
  {Manca}}, \bibinfo {author} {\bibfnamefont {M~M}\ \bibnamefont {Glazov}},
  \bibinfo {author} {\bibfnamefont {C}~\bibnamefont {Robert}}, \bibinfo
  {author} {\bibfnamefont {F}~\bibnamefont {Cadiz}}, \bibinfo {author}
  {\bibfnamefont {T}~\bibnamefont {Taniguchi}}, \bibinfo {author}
  {\bibfnamefont {K}~\bibnamefont {Watanabe}}, \bibinfo {author} {\bibfnamefont
  {E}~\bibnamefont {Courtade}}, \bibinfo {author} {\bibfnamefont
  {T}~\bibnamefont {Amand}}, \bibinfo {author} {\bibfnamefont {P}~\bibnamefont
  {Renucci}}, \bibinfo {author} {\bibfnamefont {X}~\bibnamefont {Marie}},
  \bibinfo {author} {\bibfnamefont {G}~\bibnamefont {Wang}}, \ and\ \bibinfo
  {author} {\bibfnamefont {B}~\bibnamefont {Urbaszek}},\ }\bibfield  {title}
  {\enquote {\bibinfo {title} {Enabling valley selective exciton scattering in
  monolayer wse2 through upconversion},}\ }\href@noop {} {\bibfield  {journal}
  {\bibinfo  {journal} {ArXiv}\ ,\ \bibinfo {pages} {1701.05800}} (\bibinfo
  {year} {2016})}\BibitemShut {NoStop}%
\bibitem [{\citenamefont {Taniguchi}\ and\ \citenamefont
  {Watanabe}(2007)}]{Taniguchi:2007a}%
  \BibitemOpen
  \bibfield  {author} {\bibinfo {author} {\bibfnamefont {T.}~\bibnamefont
  {Taniguchi}}\ and\ \bibinfo {author} {\bibfnamefont {K.}~\bibnamefont
  {Watanabe}},\ }\bibfield  {title} {\enquote {\bibinfo {title} {Synthesis of
  high-purity boron nitride single crystals under high pressure by using ba-bn
  solvent},}\ }\href@noop {} {\bibfield  {journal} {\bibinfo  {journal}
  {Journal of Crystal Growth}\ }\textbf {\bibinfo {volume} {303}},\ \bibinfo
  {pages} {525 -- 529} (\bibinfo {year} {2007})}\BibitemShut {NoStop}%
\bibitem [{\citenamefont {Wang}\ \emph {et~al.}(2014)\citenamefont {Wang},
  \citenamefont {Bouet}, \citenamefont {Lagarde}, \citenamefont {Vidal},
  \citenamefont {Balocchi}, \citenamefont {Amand}, \citenamefont {Marie},\ and\
  \citenamefont {Urbaszek}}]{Wang:2014b}%
  \BibitemOpen
  \bibfield  {author} {\bibinfo {author} {\bibfnamefont {G.}~\bibnamefont
  {Wang}}, \bibinfo {author} {\bibfnamefont {L.}~\bibnamefont {Bouet}},
  \bibinfo {author} {\bibfnamefont {D.}~\bibnamefont {Lagarde}}, \bibinfo
  {author} {\bibfnamefont {M.}~\bibnamefont {Vidal}}, \bibinfo {author}
  {\bibfnamefont {A.}~\bibnamefont {Balocchi}}, \bibinfo {author}
  {\bibfnamefont {T.}~\bibnamefont {Amand}}, \bibinfo {author} {\bibfnamefont
  {X.}~\bibnamefont {Marie}}, \ and\ \bibinfo {author} {\bibfnamefont
  {B.}~\bibnamefont {Urbaszek}},\ }\bibfield  {title} {\enquote {\bibinfo
  {title} {Valley dynamics probed through charged and neutral exciton emission
  in monolayer ${\mathrm{wse}}_{2}$},}\ }\href@noop {} {\bibfield  {journal}
  {\bibinfo  {journal} {Phys. Rev. B}\ }\textbf {\bibinfo {volume} {90}},\
  \bibinfo {pages} {075413} (\bibinfo {year} {2014})}\BibitemShut {NoStop}%
\bibitem [{\citenamefont {Li}\ \emph {et~al.}(2014)\citenamefont {Li},
  \citenamefont {Chernikov}, \citenamefont {Zhang}, \citenamefont {Rigosi},
  \citenamefont {Hill}, \citenamefont {van~der Zande}, \citenamefont {Chenet},
  \citenamefont {Shih}, \citenamefont {Hone},\ and\ \citenamefont
  {Heinz}}]{Li:2014b}%
  \BibitemOpen
  \bibfield  {author} {\bibinfo {author} {\bibfnamefont {Yilei}\ \bibnamefont
  {Li}}, \bibinfo {author} {\bibfnamefont {Alexey}\ \bibnamefont {Chernikov}},
  \bibinfo {author} {\bibfnamefont {Xian}\ \bibnamefont {Zhang}}, \bibinfo
  {author} {\bibfnamefont {Albert}\ \bibnamefont {Rigosi}}, \bibinfo {author}
  {\bibfnamefont {Heather~M.}\ \bibnamefont {Hill}}, \bibinfo {author}
  {\bibfnamefont {Arend~M.}\ \bibnamefont {van~der Zande}}, \bibinfo {author}
  {\bibfnamefont {Daniel~A.}\ \bibnamefont {Chenet}}, \bibinfo {author}
  {\bibfnamefont {En-Min}\ \bibnamefont {Shih}}, \bibinfo {author}
  {\bibfnamefont {James}\ \bibnamefont {Hone}}, \ and\ \bibinfo {author}
  {\bibfnamefont {Tony~F.}\ \bibnamefont {Heinz}},\ }\bibfield  {title}
  {\enquote {\bibinfo {title} {Measurement of the optical dielectric function
  of monolayer transition-metal dichalcogenides: ${\mathrm{mos}}_{2}$,
  $\mathrm{Mo}\mathrm{S}{\mathrm{e}}_{2}$, ${\mathrm{ws}}_{2}$, and
  $\mathrm{WS}{\mathrm{e}}_{2}$},}\ }\href@noop {} {\bibfield  {journal}
  {\bibinfo  {journal} {Phys. Rev. B}\ }\textbf {\bibinfo {volume} {90}},\
  \bibinfo {pages} {205422} (\bibinfo {year} {2014})}\BibitemShut {NoStop}%
\bibitem [{\citenamefont {Belhadj}\ \emph {et~al.}(2009)\citenamefont
  {Belhadj}, \citenamefont {Simon}, \citenamefont {Amand}, \citenamefont
  {Renucci}, \citenamefont {Chatel}, \citenamefont {Krebs}, \citenamefont
  {Lema\^{\i}tre}, \citenamefont {Voisin}, \citenamefont {Marie},\ and\
  \citenamefont {Urbaszek}}]{Belhadj:2009}%
  \BibitemOpen
  \bibfield  {author} {\bibinfo {author} {\bibfnamefont {Thomas}\ \bibnamefont
  {Belhadj}}, \bibinfo {author} {\bibfnamefont {Claire-Marie}\ \bibnamefont
  {Simon}}, \bibinfo {author} {\bibfnamefont {Thierry}\ \bibnamefont {Amand}},
  \bibinfo {author} {\bibfnamefont {Pierre}\ \bibnamefont {Renucci}}, \bibinfo
  {author} {\bibfnamefont {Beatrice}\ \bibnamefont {Chatel}}, \bibinfo {author}
  {\bibfnamefont {Olivier}\ \bibnamefont {Krebs}}, \bibinfo {author}
  {\bibfnamefont {Aristide}\ \bibnamefont {Lema\^{\i}tre}}, \bibinfo {author}
  {\bibfnamefont {Paul}\ \bibnamefont {Voisin}}, \bibinfo {author}
  {\bibfnamefont {Xavier}\ \bibnamefont {Marie}}, \ and\ \bibinfo {author}
  {\bibfnamefont {Bernhard}\ \bibnamefont {Urbaszek}},\ }\bibfield  {title}
  {\enquote {\bibinfo {title} {Controlling the polarization eigenstate of a
  quantum dot exciton with light},}\ }\href {\doibase
  10.1103/PhysRevLett.103.086601} {\bibfield  {journal} {\bibinfo  {journal}
  {Phys. Rev. Lett.}\ }\textbf {\bibinfo {volume} {103}},\ \bibinfo {pages}
  {086601} (\bibinfo {year} {2009})}\BibitemShut {NoStop}%
\bibitem [{\citenamefont {Jones}\ \emph {et~al.}(2015)\citenamefont {Jones},
  \citenamefont {Yu}, \citenamefont {Schaibley}, \citenamefont {Yan},
  \citenamefont {Mandrus}, \citenamefont {Taniguchi}, \citenamefont {Watanabe},
  \citenamefont {Dery}, \citenamefont {Yao},\ and\ \citenamefont
  {Xu}}]{Jones:2015a}%
  \BibitemOpen
  \bibfield  {author} {\bibinfo {author} {\bibfnamefont {Aaron~M}\ \bibnamefont
  {Jones}}, \bibinfo {author} {\bibfnamefont {Hongyi}\ \bibnamefont {Yu}},
  \bibinfo {author} {\bibfnamefont {John~R}\ \bibnamefont {Schaibley}},
  \bibinfo {author} {\bibfnamefont {Jiaqiang}\ \bibnamefont {Yan}}, \bibinfo
  {author} {\bibfnamefont {David~G}\ \bibnamefont {Mandrus}}, \bibinfo {author}
  {\bibfnamefont {Takashi}\ \bibnamefont {Taniguchi}}, \bibinfo {author}
  {\bibfnamefont {Kenji}\ \bibnamefont {Watanabe}}, \bibinfo {author}
  {\bibfnamefont {Hanan}\ \bibnamefont {Dery}}, \bibinfo {author}
  {\bibfnamefont {Wang}\ \bibnamefont {Yao}}, \ and\ \bibinfo {author}
  {\bibfnamefont {Xiaodong}\ \bibnamefont {Xu}},\ }\bibfield  {title} {\enquote
  {\bibinfo {title} {Excitonic luminescence upconversion in a two-dimensional
  semiconductor},}\ }\href@noop {} {\bibfield  {journal} {\bibinfo  {journal}
  {Nature Physics}\ } (\bibinfo {year} {2015})}\BibitemShut {NoStop}%
\bibitem [{\citenamefont {Qiu}\ \emph {et~al.}(2015)\citenamefont {Qiu},
  \citenamefont {Cao},\ and\ \citenamefont {Louie}}]{Qiu:2015a}%
  \BibitemOpen
  \bibfield  {author} {\bibinfo {author} {\bibfnamefont {Diana~Y.}\
  \bibnamefont {Qiu}}, \bibinfo {author} {\bibfnamefont {Ting}\ \bibnamefont
  {Cao}}, \ and\ \bibinfo {author} {\bibfnamefont {Steven~G.}\ \bibnamefont
  {Louie}},\ }\bibfield  {title} {\enquote {\bibinfo {title} {Nonanalyticity,
  valley quantum phases, and lightlike exciton dispersion in monolayer
  transition metal dichalcogenides: Theory and first-principles
  calculations},}\ }\href@noop {} {\bibfield  {journal} {\bibinfo  {journal}
  {Phys. Rev. Lett.}\ }\textbf {\bibinfo {volume} {115}},\ \bibinfo {pages}
  {176801} (\bibinfo {year} {2015})}\BibitemShut {NoStop}%
\bibitem [{sup()}]{suppl1}%
  \BibitemOpen
  \href@noop {} {}\bibinfo {note} {See Supplementary Material which contains
  additional experimental data and detailed group-theory and effective
  Hamiltonian analysis.}\BibitemShut {Stop}%
\bibitem [{\citenamefont {Zhang}\ \emph {et~al.}(2017)\citenamefont {Zhang},
  \citenamefont {Cao}, \citenamefont {Lu}, \citenamefont {Lin}, \citenamefont
  {Zhan}, \citenamefont {Wang}, \citenamefont {Li}, \citenamefont {Hone},
  \citenamefont {Robinson}, \citenamefont {Smirnov}, \citenamefont {Louie},\
  and\ \citenamefont {Heinz}}]{Zhang:2017a}%
  \BibitemOpen
  \bibfield  {author} {\bibinfo {author} {\bibfnamefont {X}~\bibnamefont
  {Zhang}}, \bibinfo {author} {\bibfnamefont {T}~\bibnamefont {Cao}}, \bibinfo
  {author} {\bibfnamefont {Z}~\bibnamefont {Lu}}, \bibinfo {author}
  {\bibfnamefont {Y}~\bibnamefont {Lin}}, \bibinfo {author} {\bibfnamefont
  {F}~\bibnamefont {Zhan}}, \bibinfo {author} {\bibfnamefont {Y}~\bibnamefont
  {Wang}}, \bibinfo {author} {\bibfnamefont {Z}~\bibnamefont {Li}}, \bibinfo
  {author} {\bibfnamefont {J~C}\ \bibnamefont {Hone}}, \bibinfo {author}
  {\bibfnamefont {J~A}\ \bibnamefont {Robinson}}, \bibinfo {author}
  {\bibfnamefont {D}~\bibnamefont {Smirnov}}, \bibinfo {author} {\bibfnamefont
  {S~G}\ \bibnamefont {Louie}}, \ and\ \bibinfo {author} {\bibfnamefont {T~F}\
  \bibnamefont {Heinz}},\ }\bibfield  {title} {\enquote {\bibinfo {title}
  {Magnetic brightening and control of dark excitons in monolayer wse 2},}\
  }\href@noop {} {\bibfield  {journal} {\bibinfo  {journal} {ArXiv}\ ,\
  \bibinfo {pages} {1612.03558}} (\bibinfo {year} {2017})}\BibitemShut
  {NoStop}%
\bibitem [{\citenamefont {Molas}\ \emph {et~al.}(2016)\citenamefont {Molas},
  \citenamefont {Faugeras}, \citenamefont {Slobodeniuk}, \citenamefont
  {Nogajewski}, \citenamefont {Bartos}, \citenamefont {Basko},\ and\
  \citenamefont {Potemski}}]{Molas:2017}%
  \BibitemOpen
  \bibfield  {author} {\bibinfo {author} {\bibfnamefont {M~R}\ \bibnamefont
  {Molas}}, \bibinfo {author} {\bibfnamefont {C}~\bibnamefont {Faugeras}},
  \bibinfo {author} {\bibfnamefont {A~O}\ \bibnamefont {Slobodeniuk}}, \bibinfo
  {author} {\bibfnamefont {K}~\bibnamefont {Nogajewski}}, \bibinfo {author}
  {\bibfnamefont {M}~\bibnamefont {Bartos}}, \bibinfo {author} {\bibfnamefont
  {D~M}\ \bibnamefont {Basko}}, \ and\ \bibinfo {author} {\bibfnamefont
  {M}~\bibnamefont {Potemski}},\ }\bibfield  {title} {\enquote {\bibinfo
  {title} {Brightening of dark excitons in monolayers of semiconducting
  transition metal dichalcogenides},}\ }\href@noop {} {\bibfield  {journal}
  {\bibinfo  {journal} {ArXiv:1612.02867}\ } (\bibinfo {year}
  {2016})}\BibitemShut {NoStop}%
\bibitem [{\citenamefont {{Zhou}}\ \emph {et~al.}(2017)\citenamefont {{Zhou}},
  \citenamefont {{Scuri}}, \citenamefont {{Wild}}, \citenamefont {{High}},
  \citenamefont {{Dibos}}, \citenamefont {{Jauregui}}, \citenamefont {{Shu}},
  \citenamefont {{de Greve}}, \citenamefont {{Pistunova}}, \citenamefont
  {{Joe}}, \citenamefont {{Taniguchi}}, \citenamefont {{Watanabe}},
  \citenamefont {{Kim}}, \citenamefont {{Lukin}},\ and\ \citenamefont
  {{Park}}}]{Zhou:2017}%
  \BibitemOpen
  \bibfield  {author} {\bibinfo {author} {\bibfnamefont {Y.}~\bibnamefont
  {{Zhou}}}, \bibinfo {author} {\bibfnamefont {G.}~\bibnamefont {{Scuri}}},
  \bibinfo {author} {\bibfnamefont {D.~S.}\ \bibnamefont {{Wild}}}, \bibinfo
  {author} {\bibfnamefont {A.~A.}\ \bibnamefont {{High}}}, \bibinfo {author}
  {\bibfnamefont {A.}~\bibnamefont {{Dibos}}}, \bibinfo {author} {\bibfnamefont
  {L.~A.}\ \bibnamefont {{Jauregui}}}, \bibinfo {author} {\bibfnamefont
  {C.}~\bibnamefont {{Shu}}}, \bibinfo {author} {\bibfnamefont
  {K.}~\bibnamefont {{de Greve}}}, \bibinfo {author} {\bibfnamefont
  {K.}~\bibnamefont {{Pistunova}}}, \bibinfo {author} {\bibfnamefont
  {A.}~\bibnamefont {{Joe}}}, \bibinfo {author} {\bibfnamefont
  {T.}~\bibnamefont {{Taniguchi}}}, \bibinfo {author} {\bibfnamefont
  {K.}~\bibnamefont {{Watanabe}}}, \bibinfo {author} {\bibfnamefont
  {P.}~\bibnamefont {{Kim}}}, \bibinfo {author} {\bibfnamefont {M.~D.}\
  \bibnamefont {{Lukin}}}, \ and\ \bibinfo {author} {\bibfnamefont
  {H.}~\bibnamefont {{Park}}},\ }\bibfield  {title} {\enquote {\bibinfo {title}
  {{Probing dark excitons in atomically thin semiconductors via near-field
  coupling to surface plasmon polaritons}},}\ }\href@noop {} {\bibfield
  {journal} {\bibinfo  {journal} {ArXiv e-prints}\ } (\bibinfo {year}
  {2017})},\ \Eprint {http://arxiv.org/abs/1701.05938} {arXiv:1701.05938
  [cond-mat.mes-hall]} \BibitemShut {NoStop}%
\bibitem [{Kos()}]{Koster:1963a}%
  \BibitemOpen
  \href@noop {} {}\bibinfo {note} {G. F. Koster, J. O. Dimmock, G. Wheeler, R.
  G. Satz, \textit{Properties of thirty-two point groups} (M.I.T. Press,
  Cambridge, Massachusetts USA) 1963.}\BibitemShut {Stop}%
\bibitem [{\citenamefont {Glazov}\ \emph {et~al.}(2015)\citenamefont {Glazov},
  \citenamefont {Ivchenko}, \citenamefont {Wang}, \citenamefont {Amand},
  \citenamefont {Marie}, \citenamefont {Urbaszek},\ and\ \citenamefont
  {Liu}}]{Glazov:2015a}%
  \BibitemOpen
  \bibfield  {author} {\bibinfo {author} {\bibfnamefont {M.~M.}\ \bibnamefont
  {Glazov}}, \bibinfo {author} {\bibfnamefont {E.~L.}\ \bibnamefont
  {Ivchenko}}, \bibinfo {author} {\bibfnamefont {G.}~\bibnamefont {Wang}},
  \bibinfo {author} {\bibfnamefont {T.}~\bibnamefont {Amand}}, \bibinfo
  {author} {\bibfnamefont {X.}~\bibnamefont {Marie}}, \bibinfo {author}
  {\bibfnamefont {B.}~\bibnamefont {Urbaszek}}, \ and\ \bibinfo {author}
  {\bibfnamefont {B.~L.}\ \bibnamefont {Liu}},\ }\bibfield  {title} {\enquote
  {\bibinfo {title} {Spin and valley dynamics of excitons in transition metal
  dichalcogenide monolayers},}\ }\href {\doibase 10.1002/pssb.201552211}
  {\bibfield  {journal} {\bibinfo  {journal} {physica status solidi (b)}\
  }\textbf {\bibinfo {volume} {252}},\ \bibinfo {pages} {2349--2362} (\bibinfo
  {year} {2015})}\BibitemShut {NoStop}%
\bibitem [{\citenamefont {Korn}\ \emph {et~al.}(2011)\citenamefont {Korn},
  \citenamefont {Heydrich}, \citenamefont {Hirmer}, \citenamefont
  {Schmutzler},\ and\ \citenamefont {Sch\"{u}ller}}]{Korn:2011a}%
  \BibitemOpen
  \bibfield  {author} {\bibinfo {author} {\bibfnamefont {T.}~\bibnamefont
  {Korn}}, \bibinfo {author} {\bibfnamefont {S.}~\bibnamefont {Heydrich}},
  \bibinfo {author} {\bibfnamefont {M.}~\bibnamefont {Hirmer}}, \bibinfo
  {author} {\bibfnamefont {J.}~\bibnamefont {Schmutzler}}, \ and\ \bibinfo
  {author} {\bibfnamefont {C.}~\bibnamefont {Sch\"{u}ller}},\ }\bibfield
  {title} {\enquote {\bibinfo {title} {Low-temperature photocarrier dynamics in
  monolayer ${\mathrm{mos}}_{2}$},}\ }\href@noop {} {\bibfield  {journal}
  {\bibinfo  {journal} {Applied Physics Letters}\ }\textbf {\bibinfo {volume}
  {99}},\ \bibinfo {eid} {102109} (\bibinfo {year} {2011})}\BibitemShut
  {NoStop}%
\bibitem [{\citenamefont {Lagarde}\ \emph {et~al.}(2014)\citenamefont
  {Lagarde}, \citenamefont {Bouet}, \citenamefont {Marie}, \citenamefont {Zhu},
  \citenamefont {Liu}, \citenamefont {Amand}, \citenamefont {Tan},\ and\
  \citenamefont {Urbaszek}}]{Lagarde:2014a}%
  \BibitemOpen
  \bibfield  {author} {\bibinfo {author} {\bibfnamefont {D.}~\bibnamefont
  {Lagarde}}, \bibinfo {author} {\bibfnamefont {L.}~\bibnamefont {Bouet}},
  \bibinfo {author} {\bibfnamefont {X.}~\bibnamefont {Marie}}, \bibinfo
  {author} {\bibfnamefont {C.~R.}\ \bibnamefont {Zhu}}, \bibinfo {author}
  {\bibfnamefont {B.~L.}\ \bibnamefont {Liu}}, \bibinfo {author} {\bibfnamefont
  {T.}~\bibnamefont {Amand}}, \bibinfo {author} {\bibfnamefont {P.~H.}\
  \bibnamefont {Tan}}, \ and\ \bibinfo {author} {\bibfnamefont
  {B.}~\bibnamefont {Urbaszek}},\ }\bibfield  {title} {\enquote {\bibinfo
  {title} {Carrier and polarization dynamics in monolayer mos$_2$},}\
  }\href@noop {} {\bibfield  {journal} {\bibinfo  {journal} {Phys. Rev. Lett.}\
  }\textbf {\bibinfo {volume} {112}},\ \bibinfo {pages} {047401} (\bibinfo
  {year} {2014})}\BibitemShut {NoStop}%
\bibitem [{\citenamefont {Robert}\ \emph {et~al.}(2016)\citenamefont {Robert},
  \citenamefont {Lagarde}, \citenamefont {Cadiz}, \citenamefont {Wang},
  \citenamefont {Lassagne}, \citenamefont {Amand}, \citenamefont {Balocchi},
  \citenamefont {Renucci}, \citenamefont {Tongay}, \citenamefont {Urbaszek},\
  and\ \citenamefont {Marie}}]{Robert:2016a}%
  \BibitemOpen
  \bibfield  {author} {\bibinfo {author} {\bibfnamefont {C.}~\bibnamefont
  {Robert}}, \bibinfo {author} {\bibfnamefont {D.}~\bibnamefont {Lagarde}},
  \bibinfo {author} {\bibfnamefont {F.}~\bibnamefont {Cadiz}}, \bibinfo
  {author} {\bibfnamefont {G.}~\bibnamefont {Wang}}, \bibinfo {author}
  {\bibfnamefont {B.}~\bibnamefont {Lassagne}}, \bibinfo {author}
  {\bibfnamefont {T.}~\bibnamefont {Amand}}, \bibinfo {author} {\bibfnamefont
  {A.}~\bibnamefont {Balocchi}}, \bibinfo {author} {\bibfnamefont
  {P.}~\bibnamefont {Renucci}}, \bibinfo {author} {\bibfnamefont
  {S.}~\bibnamefont {Tongay}}, \bibinfo {author} {\bibfnamefont
  {B.}~\bibnamefont {Urbaszek}}, \ and\ \bibinfo {author} {\bibfnamefont
  {X.}~\bibnamefont {Marie}},\ }\bibfield  {title} {\enquote {\bibinfo {title}
  {Exciton radiative lifetime in transition metal dichalcogenide monolayers},}\
  }\href@noop {} {\bibfield  {journal} {\bibinfo  {journal} {Phys. Rev. B}\
  }\textbf {\bibinfo {volume} {93}},\ \bibinfo {pages} {205423} (\bibinfo
  {year} {2016})}\BibitemShut {NoStop}%
\bibitem [{\citenamefont {Schardt}\ \emph {et~al.}(2006)\citenamefont
  {Schardt}, \citenamefont {Winkler}, \citenamefont {Rurimo}, \citenamefont
  {Hanson}, \citenamefont {Driscoll}, \citenamefont {Quabis}, \citenamefont
  {Malzer}, \citenamefont {Leuchs}, \citenamefont {D{\"o}hler},\ and\
  \citenamefont {Gossard}}]{Schardt:2006}%
  \BibitemOpen
  \bibfield  {author} {\bibinfo {author} {\bibfnamefont {M.}~\bibnamefont
  {Schardt}}, \bibinfo {author} {\bibfnamefont {A.}~\bibnamefont {Winkler}},
  \bibinfo {author} {\bibfnamefont {G.}~\bibnamefont {Rurimo}}, \bibinfo
  {author} {\bibfnamefont {M.}~\bibnamefont {Hanson}}, \bibinfo {author}
  {\bibfnamefont {D.}~\bibnamefont {Driscoll}}, \bibinfo {author}
  {\bibfnamefont {S.}~\bibnamefont {Quabis}}, \bibinfo {author} {\bibfnamefont
  {S.}~\bibnamefont {Malzer}}, \bibinfo {author} {\bibfnamefont
  {G.}~\bibnamefont {Leuchs}}, \bibinfo {author} {\bibfnamefont {G.H.}\
  \bibnamefont {D{\"o}hler}}, \ and\ \bibinfo {author} {\bibfnamefont {A.C.}\
  \bibnamefont {Gossard}},\ }\bibfield  {title} {\enquote {\bibinfo {title}
  {Te- and tm-polarization-resolved spectroscopy on quantum wells under normal
  incidence},}\ }\href {\doibase http://dx.doi.org/10.1016/j.physe.2005.12.168}
  {\bibfield  {journal} {\bibinfo  {journal} {Physica E: Low-dimensional
  Systems and Nanostructures}\ }\textbf {\bibinfo {volume} {32}},\ \bibinfo
  {pages} {241 -- 244} (\bibinfo {year} {2006})},\ \bibinfo {note} {proceedings
  of the 12th International Conference on Modulated Semiconductor
  StructuresProceedings of the 12th International Conference on Modulated
  Semiconductor Structures}\BibitemShut {NoStop}%
\bibitem [{\citenamefont {Bayer}\ \emph {et~al.}(2002)\citenamefont {Bayer},
  \citenamefont {Ortner}, \citenamefont {Stern}, \citenamefont {Kuther},
  \citenamefont {Gorbunov}, \citenamefont {Forchel}, \citenamefont {Hawrylak},
  \citenamefont {Fafard}, \citenamefont {Hinzer}, \citenamefont {Reinecke},
  \citenamefont {Walck}, \citenamefont {Reithmaier}, \citenamefont {Klopf},\
  and\ \citenamefont {Sch\"afer}}]{Bayer:2002}%
  \BibitemOpen
  \bibfield  {author} {\bibinfo {author} {\bibfnamefont {M.}~\bibnamefont
  {Bayer}}, \bibinfo {author} {\bibfnamefont {G.}~\bibnamefont {Ortner}},
  \bibinfo {author} {\bibfnamefont {O.}~\bibnamefont {Stern}}, \bibinfo
  {author} {\bibfnamefont {A.}~\bibnamefont {Kuther}}, \bibinfo {author}
  {\bibfnamefont {A.~A.}\ \bibnamefont {Gorbunov}}, \bibinfo {author}
  {\bibfnamefont {A.}~\bibnamefont {Forchel}}, \bibinfo {author} {\bibfnamefont
  {P.}~\bibnamefont {Hawrylak}}, \bibinfo {author} {\bibfnamefont
  {S.}~\bibnamefont {Fafard}}, \bibinfo {author} {\bibfnamefont
  {K.}~\bibnamefont {Hinzer}}, \bibinfo {author} {\bibfnamefont {T.~L.}\
  \bibnamefont {Reinecke}}, \bibinfo {author} {\bibfnamefont {S.~N.}\
  \bibnamefont {Walck}}, \bibinfo {author} {\bibfnamefont {J.~P.}\ \bibnamefont
  {Reithmaier}}, \bibinfo {author} {\bibfnamefont {F.}~\bibnamefont {Klopf}}, \
  and\ \bibinfo {author} {\bibfnamefont {F.}~\bibnamefont {Sch\"afer}},\
  }\bibfield  {title} {\enquote {\bibinfo {title} {Fine structure of neutral
  and charged excitons in self-assembled in(ga)as/(al)gaas quantum dots},}\
  }\href {\doibase 10.1103/PhysRevB.65.195315} {\bibfield  {journal} {\bibinfo
  {journal} {Phys. Rev. B}\ }\textbf {\bibinfo {volume} {65}},\ \bibinfo
  {pages} {195315} (\bibinfo {year} {2002})}\BibitemShut {NoStop}%
\end{thebibliography}

\begin{thebibliography}{6}%
\makeatletter
\providecommand \@ifxundefined [1]{%
 \@ifx{#1\undefined}
}%
\providecommand \@ifnum [1]{%
 \ifnum #1\expandafter \@firstoftwo
 \else \expandafter \@secondoftwo
 \fi
}%
\providecommand \@ifx [1]{%
 \ifx #1\expandafter \@firstoftwo
 \else \expandafter \@secondoftwo
 \fi
}%
\providecommand \natexlab [1]{#1}%
\providecommand \enquote  [1]{``#1''}%
\providecommand \bibnamefont  [1]{#1}%
\providecommand \bibfnamefont [1]{#1}%
\providecommand \citenamefont [1]{#1}%
\providecommand \href@noop [0]{\@secondoftwo}%
\providecommand \href [0]{\begingroup \@sanitize@url \@href}%
\providecommand \@href[1]{\@@startlink{#1}\@@href}%
\providecommand \@@href[1]{\endgroup#1\@@endlink}%
\providecommand \@sanitize@url [0]{\catcode `\\12\catcode `\$12\catcode
  `\&12\catcode `\#12\catcode `\^12\catcode `\_12\catcode `\%12\relax}%
\providecommand \@@startlink[1]{}%
\providecommand \@@endlink[0]{}%
\providecommand \url  [0]{\begingroup\@sanitize@url \@url }%
\providecommand \@url [1]{\endgroup\@href {#1}{\urlprefix }}%
\providecommand \urlprefix  [0]{URL }%
\providecommand \Eprint [0]{\href }%
\providecommand \doibase [0]{http://dx.doi.org/}%
\providecommand \selectlanguage [0]{\@gobble}%
\providecommand \bibinfo  [0]{\@secondoftwo}%
\providecommand \bibfield  [0]{\@secondoftwo}%
\providecommand \translation [1]{[#1]}%
\providecommand \BibitemOpen [0]{}%
\providecommand \bibitemStop [0]{}%
\providecommand \bibitemNoStop [0]{.\EOS\space}%
\providecommand \EOS [0]{\spacefactor3000\relax}%
\providecommand \BibitemShut  [1]{\csname bibitem#1\endcsname}%
\let\auto@bib@innerbib\@empty
\bibitem [{Kos()}]{Koster:1963a}%
  \BibitemOpen
  \href@noop {} {}\bibinfo {note} {G. F. Koster, J. O. Dimmock, G. Wheeler, R.
  G. Satz, \textit{Properties of thirty-two point groups} (M.I.T. Press,
  Cambridge, Massachusetts USA) 1963.}\BibitemShut {Stop}%
\bibitem [{\citenamefont {Wang}\ \emph {et~al.}(2015)\citenamefont {Wang},
  \citenamefont {Bouet}, \citenamefont {Glazov}, \citenamefont {Amand},
  \citenamefont {Ivchenko}, \citenamefont {Palleau}, \citenamefont {Marie},\
  and\ \citenamefont {Urbaszek}}]{Wang:2015d}%
  \BibitemOpen
  \bibfield  {author} {\bibinfo {author} {\bibfnamefont {G}~\bibnamefont
  {Wang}}, \bibinfo {author} {\bibfnamefont {L}~\bibnamefont {Bouet}}, \bibinfo
  {author} {\bibfnamefont {M~M}\ \bibnamefont {Glazov}}, \bibinfo {author}
  {\bibfnamefont {T}~\bibnamefont {Amand}}, \bibinfo {author} {\bibfnamefont
  {E~L}\ \bibnamefont {Ivchenko}}, \bibinfo {author} {\bibfnamefont
  {E}~\bibnamefont {Palleau}}, \bibinfo {author} {\bibfnamefont
  {X}~\bibnamefont {Marie}}, \ and\ \bibinfo {author} {\bibfnamefont
  {B}~\bibnamefont {Urbaszek}},\ }\bibfield  {title} {\enquote {\bibinfo
  {title} {Magneto-optics in transition metal diselenide monolayers},}\
  }\href@noop {} {\bibfield  {journal} {\bibinfo  {journal} {2D Materials}\
  }\textbf {\bibinfo {volume} {2}},\ \bibinfo {pages} {034002} (\bibinfo {year}
  {2015})}\BibitemShut {NoStop}%
\bibitem [{\citenamefont {Pikus}\ \emph {et~al.}(1988)\citenamefont {Pikus},
  \citenamefont {Maruschak},\ and\ \citenamefont {Titkov}}]{Pikus1988}%
  \BibitemOpen
  \bibfield  {author} {\bibinfo {author} {\bibfnamefont {G.E.}\ \bibnamefont
  {Pikus}}, \bibinfo {author} {\bibfnamefont {V.A.}\ \bibnamefont {Maruschak}},
  \ and\ \bibinfo {author} {\bibfnamefont {A.N.}\ \bibnamefont {Titkov}},\
  }\bibfield  {title} {\enquote {\bibinfo {title} {Spin splitting of
  energy-bands and spin relaxation of carriers in cubic III-V crystals},}\
  }\href@noop {} {\bibfield  {journal} {\bibinfo  {journal} {Sov. Phys.
  Semicond.}\ }\textbf {\bibinfo {volume} {22}},\ \bibinfo {pages} {115}
  (\bibinfo {year} {1988})}\BibitemShut {NoStop}%
\bibitem [{\citenamefont {{Kormanyos}}\ \emph {et~al.}(2015)\citenamefont
  {{Kormanyos}}, \citenamefont {{Burkard}}, \citenamefont {{Gmitra}},
  \citenamefont {{Fabian}}, \citenamefont {{Zolyomi}}, \citenamefont
  {{Drummond}},\ and\ \citenamefont {{Fal'ko}}}]{Kormanyos:2015a}%
  \BibitemOpen
  \bibfield  {author} {\bibinfo {author} {\bibfnamefont {A.}~\bibnamefont
  {{Kormanyos}}}, \bibinfo {author} {\bibfnamefont {G.}~\bibnamefont
  {{Burkard}}}, \bibinfo {author} {\bibfnamefont {M.}~\bibnamefont {{Gmitra}}},
  \bibinfo {author} {\bibfnamefont {J.}~\bibnamefont {{Fabian}}}, \bibinfo
  {author} {\bibfnamefont {V.}~\bibnamefont {{Zolyomi}}}, \bibinfo {author}
  {\bibfnamefont {N.~D.}\ \bibnamefont {{Drummond}}}, \ and\ \bibinfo {author}
  {\bibfnamefont {V.}~\bibnamefont {{Fal'ko}}},\ }\bibfield  {title} {\enquote
  {\bibinfo {title} {{k.p theory for two-dimensional transition metal
  dichalcogenide semiconductors}},}\ }\href@noop {} {\bibfield  {journal}
  {\bibinfo  {journal} {2D Materials}\ }\textbf {\bibinfo {volume} {2}},\
  \bibinfo {pages} {022001} (\bibinfo {year} {2015})}\BibitemShut {NoStop}%
\bibitem [{\citenamefont {Glazov}\ \emph {et~al.}(2017)\citenamefont {Glazov},
  \citenamefont {Golub}, \citenamefont {Wang}, \citenamefont {Marie},
  \citenamefont {Amand},\ and\ \citenamefont {Urbaszek}}]{PhysRevB.95.035311}%
  \BibitemOpen
  \bibfield  {author} {\bibinfo {author} {\bibfnamefont {M.~M.}\ \bibnamefont
  {Glazov}}, \bibinfo {author} {\bibfnamefont {L.~E.}\ \bibnamefont {Golub}},
  \bibinfo {author} {\bibfnamefont {G.}~\bibnamefont {Wang}}, \bibinfo {author}
  {\bibfnamefont {X.}~\bibnamefont {Marie}}, \bibinfo {author} {\bibfnamefont
  {T.}~\bibnamefont {Amand}}, \ and\ \bibinfo {author} {\bibfnamefont
  {B.}~\bibnamefont {Urbaszek}},\ }\bibfield  {title} {\enquote {\bibinfo
  {title} {Intrinsic exciton-state mixing and nonlinear optical properties in
  transition metal dichalcogenide monolayers},}\ }\href {\doibase
  10.1103/PhysRevB.95.035311} {\bibfield  {journal} {\bibinfo  {journal} {Phys.
  Rev. B}\ }\textbf {\bibinfo {volume} {95}},\ \bibinfo {pages} {035311}
  (\bibinfo {year} {2017})}\BibitemShut {NoStop}%
\bibitem [{\citenamefont {Echeverry}\ \emph {et~al.}(2016)\citenamefont
  {Echeverry}, \citenamefont {Urbaszek}, \citenamefont {Amand}, \citenamefont
  {Marie},\ and\ \citenamefont {Gerber}}]{Echeverry:2016}%
  \BibitemOpen
  \bibfield  {author} {\bibinfo {author} {\bibfnamefont {J.~P.}\ \bibnamefont
  {Echeverry}}, \bibinfo {author} {\bibfnamefont {B.}~\bibnamefont {Urbaszek}},
  \bibinfo {author} {\bibfnamefont {T.}~\bibnamefont {Amand}}, \bibinfo
  {author} {\bibfnamefont {X.}~\bibnamefont {Marie}}, \ and\ \bibinfo {author}
  {\bibfnamefont {I.~C.}\ \bibnamefont {Gerber}},\ }\bibfield  {title}
  {\enquote {\bibinfo {title} {Splitting between bright and dark excitons in
  transition metal dichalcogenide monolayers},}\ }\href {\doibase
  10.1103/PhysRevB.93.121107} {\bibfield  {journal} {\bibinfo  {journal} {Phys.
  Rev. B}\ }\textbf {\bibinfo {volume} {93}},\ \bibinfo {pages} {121107}
  (\bibinfo {year} {2016})}\BibitemShut {NoStop}%
\end{thebibliography}

%

\end{document}